\documentclass[12pt]{iopart}

\usepackage{iopams}
\expandafter\let\csname equation*\endcsname\relax
\expandafter\let\csname endequation*\endcsname\relax
\usepackage[usenames,dvipsnames]{xcolor}
\colorlet{color1}{NavyBlue}
\usepackage[colorlinks=true,allcolors=color1]{hyperref}
\usepackage{amssymb}
\usepackage{physics}
\usepackage{dsfont}
\usepackage{graphicx}
\usepackage{xfrac,overpic}
\usepackage[utf8]{inputenc}

\usepackage{soul}

\begin{document}
\title{Semiclassical gravity effects near horizon formation}

\author{Carlos Barceló$^1$, Valentin Boyanov$^{2}$, Raúl Carballo-Rubio$^{3,4,5}$ and Luis J. Garay$^{2,6}$}

\address{$^1$Instituto de Astrofísica de Andalucía (IAA-CSIC), Glorieta de la Astronomía, 18008 Granada, Spain\\
$^2$Departamento de Física Teórica and IPARCOS, Universidad Complutense de Madrid, 28040 Madrid, Spain\\
$^3$SISSA - International School for Advanced Studies, Via Bonomea 265, 34136 Trieste, Italy\\
$^4$IFPU - Institute for Fundamental Physics of the Universe, Via Beirut 2, 34014 Trieste, Italy\\
$^5$INFN Sezione di Trieste, Via Valerio 2, 34127 Trieste, Italy\\
$^6$Instituto de Estructura de la Materia (IEM-CSIC), Serrano 121, 28006 Madrid, Spain}
\ead{carlos@iaa.es, vboyanov@ucm.es, raul.carballorubio@sissa.it and luisj.garay@ucm.es}
\vspace{10pt}
\begin{indented}
	\item[]April 2019
\end{indented}

\begin{abstract}
	We study the magnitude of semiclassical gravity effects near the formation of a black-hole horizon in spherically-symmetric spacetimes. As a probe for these effects we use a quantised massless scalar field. Specifically, we calculate two quantities derived from it: the renormalised stress-energy tensor (a measure of how the field vacuum state affects the spacetime) and the effective temperature function (a generalisation of Hawking temperature related to the energy flux of the field vacuum). The subject of our study are spacetimes which contain a spherical distribution of matter and an empty exterior Schwarzschild region, separated by a surface which is moving in proximity to the Schwarzschild radius $r_{\rm s}=2M$, with $M$ the total mass. In particular, we analyse the consequences of three types of surface movement: an oscillation just above $r_{\rm s}$, a monotonous approach towards $r_{\rm s}$ in infinite time and a crossing of $r_{\rm s}$ at different velocities. For a collapsing matter distribution which follows the expected dynamical evolution in general relativity, we recover the standard picture of black-hole formation and its tenuous semiclassical effects. In more general dynamical regimes, allowing deviations from the standard classical evolution, we obtain a variety of different effects: from the emission of Hawking-like radiation without the formation of a horizon, to large values of the renormalised stress-energy tensor, related to the Boulware vacuum divergence in static spacetimes.
\end{abstract}

\maketitle

\section{Introduction}
The theory of quantum fields in curved spacetimes remains, to this day, one of the leading methods for studying quantum corrections to the evolution of the spacetime geometry. Historically, one of the most notable results originating from this theory is that of Hawking radiation~\cite{Hawking1975}: the formation of a black hole entails the emission of thermal radiation, which through back-reaction can translate into the slow depletion of the black-hole mass. This quantum effect was later also shown to occur in the presence of cosmological horizons in an expanding universe~\cite{GH1977}.\par
The semiclassical theory is characterised by keeping a classical spacetime background framework, but introducing quantum corrections to the energy momentum density which governs its dynamics. This is achieved through calculating the renormalised stress-energy tensor (RSET) of quantum fields, which takes the stress-energy content of the quantum vacuum to the same footing as the classical matter term in the Einstein equations. This semiclassical approach can be considered an intermediate step in the way toward a quantum theory of gravity, in which the interactions of matter and spacetime would reveal their quantum characteristics more naturally and fully (see e.g. \cite{BD} for a discussion on its interpretation as an approximation to quantum gravity). The general consensus is that this approximation is valid in regions of spacetime with sufficiently low, non-Planckian curvature (e.g. the horizon of astrophysical black holes), where it provides insightful information about quantum corrections to classical general relativity. These corrections are, however, suppressed by Planck's constant, and although they have far-reaching conceptual implications, they appear practically irrelevant for most astrophysical processes. Still, there might exist situations in which this suppression could be overcome, as we will see in this work.\par
When quantising a field in a curved spacetime, there is no preferred slicing in spacelike hypersurfaces (orthogonal to a timelike vector field) on which to define an operator algebra. This leads to a corresponding ambiguity in the choice of the vacuum and particle states of the field. An interesting example which illustrates this ambiguity, and is particularly relevant to this work, is the following. If we consider a static black hole and quantise with respect to the Schwarzschild time coordinate, we get a definition of particles which becomes more similar to the one in Minkowski spacetime the farther away from the horizon you are, matching perfectly in the asymptotic region where spacetime is flat (the corresponding vacuum state is known as the Boulware vacuum \cite{Candelas1980,Boulware}). However, at the horizon itself operator expectation values in this state present a divergent behaviour, owing to the irregularity of the Schwarzschild time coordinate there. This particular quantisation is therefore deemed as nonphysical, backing up the idea that there should not exist \textit{eternal} black holes in our universe which would require it. A somewhat more physically reasonable scenario is that of an asymptotically flat (ignoring cosmological backgrounds) spacetime with an initially dispersed distribution of matter, which eventually collapses to form a black hole. In it, we can choose an asymptotically Minkowskian quantisation in the asymptotic past, which can be extended to the whole spacetime (through the solutions of the field equation). The vacuum state for this quantisation is known as the $in$ vacuum, for which the Hawking radiation result was obtained, and in which, unlike the case of the Boulware vacuum state, observables are regular at the horizon. The $in$ vacuum is in fact the physical vacuum state one should consider when dealing with a black-hole formation process in an asymptotically flat spacetime.\par
When using this vacuum, the overall resulting picture is that any stellar-mass object which \emph{collapses rapidly toward the formation of a horizon} generates extremely small RSETs. It is important to stress that ``rapidly'' in the previous sentence corresponds precisely to the standard situation one would expect when working in the framework of general relativity (defined by the Einstein field equations coupled to matter satisfying the standard energy conditions \cite{Barcelo2002,Curiel2014}) and taking into account the forces that are known to play a role in stellar evolution. In these situations, semiclassical effects are in fact so small, that the collapse would proceed in almost exactly the same manner as in classical general relativity, forming a trapping horizon and continuing until the appearance of a Planck curvature region (see e.g. \cite{DFU,Parentani1994} for the first treatments of this problem and \cite{Barcelo2008,Unruh2018} for modern retakes). The crucial hypothesis of ``\emph{rapid approach toward the formation of a horizon}" is, therefore, perfectly sensible in most scenarios. However, the presence of a quantum bounce at Planck curvatures \cite{Barcelo2014,Barcelo2014b,Haggard2015} or the presence of metastable states before horizon formation \cite{Barcelo2008} might lead to situations in which this hypothesis is questionable. For instance, the divergent behaviour of the Boulware vacuum may be taken as a hint of the possibility that even in a physical vacuum, the surroundings of a black-hole horizon may be a region where semiclassical corrections become large enough to be relevant to the evolution of the system. Indeed, as we will show, the hypothetical formation of ultracompact objects sustained very close to horizon formation (an alternative to black holes \cite{Visser2004,Visser2008,Cardoso2017,Cardoso2019,Carballo-Rubio2018}) appears to require at least a semiclassical treatment. Generally, if the RSET contribution overcomes its suppression by Planck's constant and becomes comparable to the classical stress-energy tensor, then a complete, non-perturbative semiclassical treatment of the problem is in order.\par
In this work we study the values of the RSET for the $in$ vacuum of a free massless scalar field in spherically-symmetric geometries which approach the formation of a horizon in different ways. Previous works with the same motivation have checked some of the semiclassical effects produced by a collapse of matter which quickly decelerates just before reaching the formation of a horizon \cite{Pad2009,Harada2019}. Our present goal is to more generally identify the precise geometric characteristics of the dynamical situations which would cause large back-reaction close to horizon formation.\par
In section \ref{s2} we start by reviewing the definitions of the functions which measure the deviation from classical physics. One of them is the already mentioned RSET, which directly serves as a source in the Einstein equations. The other is the effective temperature function (ETF) introduced in \cite{BLSV11}. This is a generalisation of the Hawking temperature which characterises the flux of outgoing radiation at future infinity. As was shown in \cite{BBGJ16}, this function is directly related to the term in the RSET evaluated in dynamical vacua which regularises the divergence at the horizon of the static Boulware vacuum.\par
Also in sec. \ref{s2}, we will introduce the generic structure of the geometries we will use, namely a Schwarzschild exterior and a Minkowski interior, separated by a thin spherical shell which moves radially along some timelike curve. We will provide a physical interpretation for the relations between the static null coordinates corresponding to the interior and exterior regions, which will allow us to understand how the modes of the massless field are dispersed when crossing the shell. After this, we will move on to specifying what types of trajectories for the shell will be used: oscillations close to (but above) the horizon, asymptotic approach to horizon formation and actual formation of a horizon at low velocities. These are the three types of curves which exploit the physics of being close to horizon formation. Before embarking on a detailed study of these cases, we will explain briefly the different notions of horizon that one can define and their relevance for our analysis. In this introduction we have been deliberately vague in this respect; let us only advance that ``horizon'' must be identified with the notion of apparent/trapping horizon and not with event horizon, unless explicitly stated. Let us also note that most of the results in this paper will be phrased in terms of an exterior Schwarzschild geometry, but they are in fact more general and will apply equally to any exterior geometry with a non-zero surface gravity at the horizon.\par
In section \ref{s4} we will study shell trajectories which oscillate just above the Schwarzschild radius $r_{\rm s}=2M$, $M$ being the mass of the shell (note that we will always be using natural units $G=\hbar=c=1$). This will serve as an example which illustrates how the relations between the null coordinates are associated to semiclassical effects in different dynamical regimes close to horizon formation.\par
In section \ref{s5} we will explore shell trajectories which approach the horizon so slowly that they do not reach it in finite time. For this case, as we will show, the thin-shell approximation misses some important effects. Therefore we will extend our study to an arbitrary spherically-symmetric geometry for the interior region. The results we are interested in will be obtained through a study of the asymptotic future values of the ETF, which will allow us to compare the $in$ vacuum state to the Boulware vacuum.\par
In section \ref{s6} we will go back to the thin-shell approximation and study geometries in which a horizon is formed in finite time. In this case the asymptotic behaviour of both the ETF and the RSET are well-known \cite{Hawking1975,DFU} so we will focus on their values at times close to the formation of the horizon. Particularly, we will vary the velocity at which the shell crosses the $r=2M$ surface, paying close attention to the lower velocity results. In the final section we will summarise our findings.

\section{Preliminaries}\label{s2}

The overall aim of this work is to gauge the magnitude of semiclassical effects in a series of specific dynamical and spherically-symmetric geometries, characterised for being close to the formation of a horizon. As is common in black-hole physics, we will use a single massless scalar field as a probe. For the calculation of the RSET, we will be making use of an analytic approximation to its exact form, which amounts to considering only the $s$-wave contributions and neglecting the backscattering effects of the geometry. As we will see, the closeness of the spacetime to the formation of a horizon is going to be a key factor for the behaviour of the RSET.

\subsection{Renormalised stress-energy tensor in 1+1 dimensions}

The analytic approximation to the RSET we will be considering takes advantage of the conformal invariance of a massless scalar field equation in 1+1 dimensions. All 4-dimensional spherically-symmetric metrics can be written as
\begin{equation}\label{21}
ds^2=-C(u,v)dudv+r^2d\Omega_2^2,
\end{equation}
where $C(u,v)$ is a positive function of the radial null coordinates (non-zero for regular coordinates). The calculation of the dimensionally reduced 1+1 RSET attends exclusively to the $\{u,v\}$ coordinates, i.e. to the radial-temporal part of the geometry. In this approximation, $C(u,v)$ is the conformal factor which rescales the otherwise flat two-dimensional spacetime.\par
A massless scalar field $\phi$ can be Fock-quantised on this background with a basis of solutions of the Klein-Gordon equation given by the ingoing and outgoing modes
\begin{equation}\label{22}
\phi_\omega^u=\frac{1}{\sqrt{4\pi\omega}}e^{-i\omega u},\quad \phi_\omega^v=\frac{1}{\sqrt{4\pi\omega}}e^{-i\omega v},
\end{equation}
where $\omega>0$. Such a basis exists for any pair of null coordinates which cover the spacetime. Choosing a particular pair amounts to a choice of vacuum and particle states of the quantum field.\par
The RSET is the renormalised vacuum expectation value of the stress-energy tensor operator constructed with the quantum field. For a choice of null coordinates $\{u,v\}$ and a corresponding vacuum $\ket{0}$, the components of the RSET in this same coordinate basis are (see \cite{DF})
\begin{subequations}\label{23}
	\begin{align}
	\expval{T_{uu}}_0&=\frac{1}{24\pi}\left[\frac{\partial_u^2C}{C}-\frac{3}{2}\left(\frac{\partial_uC}{C}\right)^2\right],\\
	\expval{T_{vv}}_0&=\frac{1}{24\pi}\left[\frac{\partial_v^2C}{C}-\frac{3}{2}\left(\frac{\partial_vC}{C}\right)^2\right],\\
	\expval{T_{uv}}_0&=\frac{1}{24\pi}\left[\frac{\partial_uC\partial_vC}{C^2}-\frac{\partial_u\partial_vC}{C}\right].
	\end{align}
\end{subequations}
Only the trace of the RSET depends explicitly on the curvature: $\expval{T_\mu^\mu}=R/24\pi$ (see e.g. \cite{BD,DF,FN}). On the other hand, the traceless part of the RSET can be expressed in a geometric form in terms of the norm of the timelike vector field $\partial_t=\partial_u+\partial_v$ that defines the vacuum state $|0\rangle$ \cite{Barcelo2011}. This means that the traceless (and state-dependent) part of the RSET could in principle become large in regions of low curvature.\par
The choice of quantisation modes and vacuum state for this theory is arbitrary. In a different vacuum state $\ket{\tilde{0}}$, corresponding to a quantisation characterised by a different pair of coordinates $\{\tilde{u},\tilde{v}\}$, related to the first through two positive functions $g$ and $h$ such that
\begin{equation}\label{24}
\frac{du}{d\tilde{u}}=g(\tilde{u}),\quad \frac{dv}{d\tilde{v}}=h(\tilde{v}),
\end{equation}
the components of the RSET in the first $\{u,v\}$ coordinate basis are related to those in the $\ket{0}$ vacuum through
\begin{subequations}\label{25}
	\begin{align}
	\expval{T_{uu}}_{\tilde{0}}&=\frac{1}{24\pi}\left(\frac{g''}{g^3}-\frac{3}{2}\frac{g'^2}{g^4}\right)+\expval{T_{uu}}_0,\\
	\expval{T_{vv}}_{\tilde{0}}&=\frac{1}{24\pi}\left(\frac{h''}{h^3}-\frac{3}{2}\frac{h'^2}{h^4}\right)+\expval{T_{vv}}_0,\\
	\expval{T_{uv}}_{\tilde{0}}&=\expval{T_{uv}}_0,
	\end{align}
\end{subequations}
where $g'\equiv\partial_{\tilde{u}}g(\tilde{u})$ and $h'\equiv\partial_{\tilde{v}}h(\tilde{v})$. With these expressions we can see that a change in the vacuum state translates into the addition of outgoing and ingoing radiation flux terms (which can be positive or negative).\par
Apart from the RSET, we are interested in studying the values of the effective temperature function (ETF) \cite{BLSV11}, defined as
\begin{equation}\label{26}
\kappa_{\tilde{u}}^u\equiv-\left.\frac{d^2\tilde{u}}{du^2}\right/\frac{d\tilde{u}}{du}=\frac{g'}{g^2}
\end{equation}
for the outgoing radiation sector, and likewise substituting $u$'s for $v$'s (and $g$ for $h$) for the ingoing sector. In the case of a spacetime representing the formation of a black hole, the usual Hawking effect is reflected in the constant value $\kappa_{u_{in}}^{u_{out}}=1/2=2\pi T_{\rm H},$ where $T_{\rm H}$ is the Hawking temperature in natural units. In more general terms, if $\kappa^u_{\tilde{u}}$ or $\kappa^v_{\tilde{v}}$ remain constant for a sufficiently long period of time (defined by an adiabaticity condition), the vacuum state defined by the $\{\tilde{u},\tilde{v}\}$ coordinates (through the modes in~\eqref{22}) will be seen by an observer with proper coordinates $\{u,v\}$ as a thermal state of outgoing or ingoing radiation respectively \cite{BLSV11}.\par
This function is also directly related to the outgoing and ingoing radiation fluxes which appear in the RSET after a change of vacuum state \cite{BBGJ16}. Specifically, equations~\eqref{25} can be written as
\begin{subequations}\label{27}
	\begin{align}
	\expval{T_{uu}}_{\tilde{0}}&=\frac{1}{24\pi}\left(\frac{d\kappa_{\tilde{u}}^u}{du}+\frac{1}{2}(\kappa_{\tilde{u}}^u)^2\right)+\expval{T_{uu}}_0,\\
	\expval{T_{vv}}_{\tilde{0}}&=\frac{1}{24\pi}\left(\frac{d\kappa_{\tilde{v}}^v}{dv}+\frac{1}{2}(\kappa_{\tilde{v}}^v)^2\right)+\expval{T_{vv}}_0,\\
	\expval{T_{uv}}_{\tilde{0}}&=\expval{T_{uv}}_0.
	\end{align}
\end{subequations}
In other words, the information about the difference between the RSETs in two different vacuum states is entirely contained in their relative ETFs (and first derivatives thereof).\par
When calculating either of these quantities, the information about the choice of vacuum state is encoded in specific sets of null coordinates. The fact that the RSET is defined by relations between null coordinates is the reason why its expression is not generally given by geometric quantities that can be reduced to curvature invariants, and why it can in principle become large in regions of low curvature (as we will see explicitly). For the spacetimes we will study, we are interested in calculating these quantities for two special quantum vacuum states: the $in$ and the $out$ states. The $in$-state ($out$-state) is the one defined by affine null coordinates at past (future) null infinity. In order to carry out the calculation, we will want to extend these sets of coordinates throughout the whole spacetime, if possible, and obtain the relations between them. However, if there is a horizon present at some point, one or both of these extensions may cover the spacetime only partially. For example, in a collapsing geometry which starts being almost flat and ends up forming a black hole, the $in$-state corresponds to the natural Minkowski vacuum at the asymptotic past which then evolves according to the dynamics of the system. On the other hand, the $out$-state would correspond asymptotically to the Boulware state, and its extension backward in time would cover only the region of spacetime outside the event horizon.

\subsection{Thin-shell geometries with spherical symmetry}
The geometries that we will analyse all consist of an internal Minkowskian region pasted to an external Schwarzschild region of mass $M$ through a moving timelike shell. In the interior region one can write the metric as
\begin{equation}\label{1}
ds_-^2=-du_-dv_-+r_-^2d\Omega^2,
\end{equation}
where the subscript ``$-$" refers to the interior region, and the radial null coordinates are related to the Minkowski time $t_-$ and radius $r_-$ through
\begin{equation}\label{2}
	u_-=t_--r_-,\quad v_-=t_-+r_-.
\end{equation}
Equivalently we can construct natural null coordinates in the Schwarzschild region as
\begin{equation}\label{3}
ds_+^2=-|f(r_+)|du_+dv_++r_+^2d\Omega^2,
\end{equation}
where $f(r)=1-2M/r$ is the redshift function, and in this case the null coordinates are related to the Schwarzschild time $t_+$ and radius $r_+$ through
\begin{equation}\label{eq:u+def}
	u_+=\text{sign}\left[f(r_+)\right](t_+-r_+^*),\quad v_+=t_++r_+^*.
\end{equation}
Here $r_+^*$ is the tortoise coordinate obtained by integrating $dr_+^*=dr_+/f(r_+)$. The $u_+$ coordinate goes from $-\infty$ to $+\infty$, that is, between past null infinity and the Schwarzschild radius (if the exterior region reaches that far in). Inside the Schwarzschild radius (but outside the shell) we must define a different coordinate $u^i_+$, given by the same relation to $t_+$ and $r_+^*$ as $u_+$ above, and which goes from $-\infty$ at the horizon, until it reaches some point of the spacelike singularity at some finite value. On the horizon itself, relations with this variable can only be obtained as a limit from either side. The sign of $f(r_+)$ ensures that $u_+$ and $u^i_+$ advance in the same direction as $u_-$, both outside and inside the horizon.\par
The two geometries are connected by a thin spherical shell of mass $M$. In general, this matching is only possible if the shell's radial position follows a spacetime curve of the same causality type as seen from either side. In our case, we will require that this be a timelike trajectory, parametrised by $v_-=T_-(u_-)$ from the inside and by $v_+=T_+(u_+)$ from the outside. Of course, given one of these curves the other is also fixed. For convenience we will also define the velocity parameters
\begin{equation}\label{5}
\alpha_-\equiv\left.\frac{dv_-}{du_-}\right|_{\rm shell},\qquad \alpha_+\equiv\left.\frac{dv_+}{du_+}\right|_{\rm shell}
\end{equation}
(which are simply the derivatives of $T_\pm$), both of which take values in $(0,\infty)$ for a timelike trajectory. For an ingoing shell to approach the speed of light would imply approaching the limit $\alpha_\pm\to0$. On the other hand, for an outgoing shell reaching the speed of light $\alpha_{\pm}\to\infty$. A static shell has $\alpha_\pm=1$.\par
In order to complete the definition of this geometry, we must require that the metric be continuous at the shell. This will allow us to determine the trajectory of the shell as seen from one side if it is defined on the other. It will also allow us to extend the ``$+$" coordinates into the ``$-$" region and vice versa.\par
From matching the null part of the line elements we obtain the functions defined in \eqref{24},
\begin{equation}\label{6}
g=\frac{du_+}{du_-}=\left.\sqrt{\frac{\alpha_-}{|f|\alpha_+}}\right|_{\rm shell},\qquad h=\frac{dv_+}{dv_-}=\left.\sqrt{\frac{\alpha_+}{|f|\alpha_-}}\right|_{\rm shell},
\end{equation}
which can be expressed in either the ``$+$" or ``$-$" variables. From matching the radial parts we get the relation between the velocity parameters of the shell from either side,
\begin{equation}\label{7}
\alpha_+=\text{sign}(f)+\frac{1}{2|f|}\frac{(1-\alpha_-)^2}{\alpha_-}-\frac{1}{2|f|}\frac{1-\alpha_-}{\alpha_-}\sqrt{4\alpha_- f+(1-\alpha_-)^2}.
\end{equation}
Thus if we define the trajectory in terms of $T_-$, we can obtain $T_+$ by integrating $\alpha_+$ from the same initial radial position. We can also obtain the relations $u_+(u_-)$ and $v_+(v_-)$ by integrating the functions $g$ and $h$.\par
From the square root in \eqref{7} we deduce a condition for the continuous matching of the geometries, namely that the $\alpha_-$ parameter which defines the movement of their separation surface must be such that $4\alpha_- f+(1-\alpha_-)^2$ remains positive. In other words, $\alpha_-$ must tend to zero (the infalling shell must approach light-speed) inside the Schwarzschild radius in such a way as to compensate the increasingly negative value of the redshift function. The parameter $\alpha_-$ which satisfies
\begin{equation}
4\alpha_- f|_{\rm shell}+(1-\alpha_-)^2=0
\end{equation}
defines the slowest possible collapse inside the event horizon as seen from the (rapidly disappearing) Minkowski region.

\subsection{Interpretation of the terms in $g$ and $h$}
Let us focus on the function $g$ outside the Schwarzschild radius,
\begin{equation}\label{8}
g=\frac{du_+}{du_-}=\frac{1}{\sqrt{f}}\sqrt{\frac{\alpha_-}{\alpha_+}}.
\end{equation}
The presence of the term $1/\sqrt{f}$ is to be expected, as it represents the redshift experienced by an outgoing light ray. This can be seen most clearly in the case of a static shell (which, of course, would sit outside the horizon), for which $\alpha_\pm=1$. There, it is necessary for a rescaling of the coordinates compatible with a matching of the angular parts of the geometry.\par
\begin{figure}
	\centering
	\includegraphics[scale=.7]{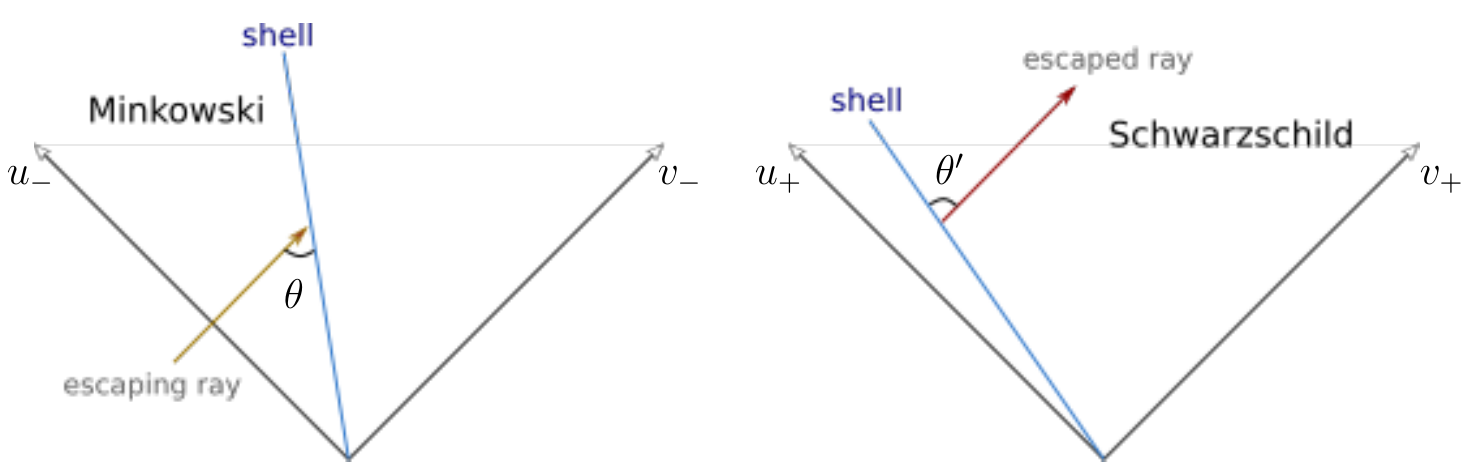}
	\caption{Change in angle with respect to the shell of on outgoing light ray, as measured by static observers on either side.}
	\label{f1}
\end{figure}

The $\alpha_-/\alpha_+$ term has a purely dynamical origin. The velocity of the shell seen by a static observer on one of its sides is different from the one seen by a static observer on the other. In their respective null coordinates this can be seen as a change in the slope of the line tangent to the shell trajectory, namely $\alpha_-\to\alpha_+$ (see fig.~\ref{f1}). From the perspective of the shell, which can use the appropriate coordinates for each side, this looks something like a spacetime refraction phenomenon. If a light ray incides with an angle $\theta$ with respect to the shell trajectory from the inside, it exits with an angle $\theta'$ related to the first by
\begin{equation}
\frac{\tan \theta'}{\tan \theta}=\frac{\alpha_-}{\alpha_+}.
\end{equation}
For the angles formed by an ingoing ray, the relation is the inverse of the above.
\par
Another way to interpret the $\alpha_-/\alpha_+$ term is as a kind of Doppler effect, even though technically there is no interaction between the matter in the shell and the light ray crossing it which could cause absorption and reemission. The effect can be seen clearly with the following. If we define
\begin{equation}
R(t_\pm)\equiv r|_{\rm shell}(t_\pm)\quad\text{and}\quad \dot{R}=dR/dt_-,\quad R'=dR/dt_+,
\end{equation}
then
\begin{equation}\label{10}
	\alpha_-=\frac{1+\dot{R}}{1-\dot{R}},\quad \alpha_+=\frac{1+R'/f(R)}{1-R'/f(R)}. 
\end{equation} 
That is, the quotient $\alpha_-/\alpha_+$ represents the Doppler shift for a ray that is ``absorbed" at one side by a shell moving at a velocity $\dot{R}$ and ``reemitted" on the other by a shell moving at a different velocity, $R'/f(R)$. If the geometry on both sides were the same, there would be no net effect, as these velocities would be the same.\par
It is worth mentioning that there may be a difficulty in interpreting the above expressions at the Schwarzschild radius, since the coordinate $t_+$ used for the derivative in the second equation in \eqref{10} is not regular there. To see the behaviour of $\alpha_+$ more clearly we can switch to a regular time coordinate, say the Painlevé-Gullstrand $\tau_+$ defined as the proper time of a free-falling observer from infinity in the Schwarzschild region \cite{Martel2001}, which satisfies
\begin{equation}
d\tau_+=dt_++\frac{\sqrt{1-f(r_+)}}{f(r_+)}dr_+.
\end{equation}
We can then define the radial velocity $R_{,\tau}\equiv dR/d\tau_+$, which is regular at the horizon. Then the second equation in \eqref{10} becomes
\begin{equation}\label{12}
\alpha_+=\frac{1+R_{,\tau}/(1+\sqrt{2M/R})}{1-R_{,\tau}/(1-\sqrt{2M/R})},
\end{equation}
from which we can see that at $r=2M$, $\alpha_+=0$ and the function $g$ in \eqref{8} diverges.\par
In light of these results, we will call the $1/\sqrt{|f|}$ terms in the functions $g$ and $h$ the ``redshift" terms, and the ones with a quotient of $\alpha$'s the ``Doppler" terms. Combining equations \eqref{6}, \eqref{10} and \eqref{12} we can write the total functions as
\begin{subequations}\label{13}
	\begin{align}
	&g(u)=\frac{1}{\sqrt{f}}\sqrt{\frac{1+\dot{R}}{1-\dot{R}}}\sqrt{\frac{1-R_{,\tau}/(1-\sqrt{2M/R})}{1+R_{,\tau}/(1+\sqrt{2M/R})}},\\
	&h(v)=\frac{1}{\sqrt{f}}\sqrt{\frac{1-\dot{R}}{1+\dot{R}}}\sqrt{\frac{1+R_{,\tau}/(1+\sqrt{2M/R})}{1-R_{,\tau}/(1-\sqrt{2M/R})}},
	\end{align}
\end{subequations}
in which all quantities are evaluated at the points where the lines $u=const.$ and $v=const.$ intersect the shell trajectory respectively. We could work directly with these expressions instead of \eqref{6} by defining the trajectory through the velocities $\dot{R}$ and $R_{,\tau}$, which must satisfy a relation similar to \eqref{7}. However, throughout this work we will keep using the $\alpha_\pm$ parameters, as they are more natural and simple when dealing with the relations between null coordinates needed for the calculation of semiclassical quantities.

\subsection{Geometries near horizon formation}
The geometries we are going to study are all characterised by being close to horizon formation in a specific sense that is clarified in this section. Probably the best-known characterisation of black holes is the classical one in terms of event horizons, the definition of which exploits the notion of future null infinity in asymptotically flat spacetimes \cite{Hawking1973}. However, event horizons display a number of undesirable features, such as their lack of univocal relation with strong gravitational fields \cite{Ashtekar2004}, or their nonlocal nature that forbids their detection in experiments (that necessarily take place in finite regions of spacetime) \cite{Visser2014}. Even if the geometries analysed below contain event horizons, the notion of being close to horizon formation that is relevant to our analysis is always described in terms of local properties of spacetime and matter fields, and is closely related to (quasi-)local definitions of the boundaries of black holes in terms of apparent horizons. One can alternatively use the concept of trapping horizons or other equivalent definitions (see \cite{Gourgoulhon2008}, for instance, for a review); however, in the situations analysed here, all these definitions become equivalent, so there is no need for us to discuss their differences. In some of the geometries analysed below, the position of apparent/trapping horizons and event horizons are coincident. This should not be taken as an indication that our results are tied in any way with the formation of event horizons. In fact, it is always possible to deform these geometries in a way that event horizons are removed completely, but the local geometric conditions that eventually lead to their formation in the undeformed geometries are maintained for arbitrarily long times (for geometries in which apparent/trapping horizons are formed in finite time, this would imply that they remain present for a large, but finite, amount of time), which would yield the same results but for arbitrarily small deviations.\par
One of the shortcomings of these (quasi-)local definitions of the boundaries of black holes (with respect to the notion of event horizon) is their non-uniqueness \cite{Ashtekar2005}. This issue disappears in practice when dealing with spherically-symmetric backgrounds, as one can focus on trapping horizons that are spherically-symmetric as well, the location of which turn out to be determined by the quasi-local Misner-Sharp mass \cite{Nielsen2008} that measures the overall energy enclosed in a given sphere \cite{Hayward1994}. When the external geometry to the shell is the Schwarzschild geometry, the location of the horizon defined this way is simply the Schwarzschild radius.\par
In this work, ``close to horizon formation'' will therefore mean that the shell has trajectories exploring the surroundings of the Schwarzschild radius. There, we expect to find interesting semiclassical effects, and we want to understand their dependence on the precise dynamical properties of the spacetime as it approaches this point. To this end, we have chosen three types of shell trajectories, the study of which we believe will lead to the necessary insight for any general situation. The first type of situation, studied in sec. \ref{s4}, is when a shell oscillates between two radii, outside but near the Schwarzschild radius. This situation models in the easiest terms ultracompact objects subject to small pulsations. Varying the characteristics of this oscillation will allow us to explore a wide range of short-term dynamical behaviours. The second type of situation (sec. \ref{s5}) will explore the consequences of a long-term monotonous dynamical behaviour, particularly one which we expect (both \textit{a priori} and based on results of sec. \ref{s4}) to present interesting semiclassical effects -- a shell approaching the Schwarzschild radius asymptotically in a regular time coordinate. As we will see, in this study it will become apparent that we need to go beyond the thin-shell approximation. The asymptotic approach can be stopped at any time, so that these configurations could model, for example, a relaxation phase towards an ultracompact object. For the third type we go back to thin shells, and conclude our study with the case in which they actually form a horizon in finite regular time, though they do so while moving at an arbitrarily slow pace. Our analysis here, which is an extension to \cite{Barcelo2008}, allows us to clearly see how the strength of semiclassical effects depends crucially on the collapsing velocity at horizon formation. This will provide a counterpoint to the already well-known results for a shell collapsing at high velocities or even light speed (see e.g. \cite{FN}).\par
It is worth mentioning at this point that our analysis throughout this work will be purely geometrical, and thus goes beyond the Einstein equations. In other words, we will be exploring the effects of a geometry on semiclassical quantities without being concerned with how the geometry itself is generated. In other words, we will not require that the evolution of the geometry be governed by the Einstein equation with a stress-energy tensor which satisfies some energy conditions. Although it is certainly interesting to study the properties of the matter content (both classical and semiclassical) which would generate the geometries in question, this lies beyond the scope of the present work. Instead, our geometry-based results will just point the way toward the configurations which should be analysed in further detail in future works in the context of semiclassical gravity.

\section{Oscillating thin shells}\label{s4}

\begin{figure}
	\centering
	\includegraphics[scale=.6]{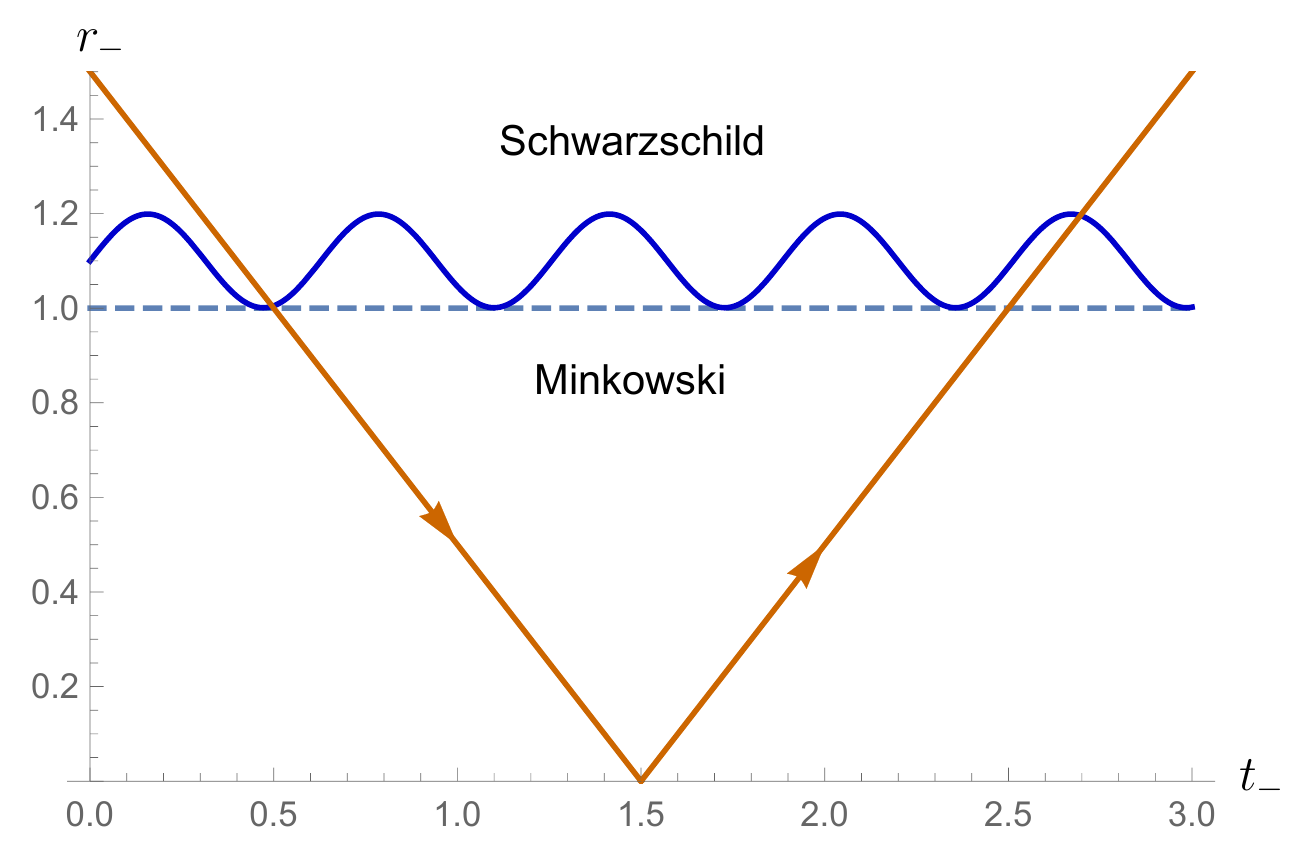}	
	\caption{Oscillatory radial trajectory of the shell (with parameters $d=0.1$, $A=0.099$ and $\omega=10$). The dashed line represents the $r=2M$ $(=1)$ surface, and the diagonal lines represent a light ray entering and exiting the interior region. Although not perceived in the figure, the thick oscillatory curve does not touch the $r=2M$ line.}
	\label{f2}
\end{figure}

In this section we will study the behaviour of the functions $g$ and $h$, which relate the ``$+$" and ``$-$" coordinates, when the shell gets near the formation of a horizon, but does not reach it. Nonetheless, it will follow a trajectory which covers a wide range of dynamical configurations, in which both the redshift and Doppler effects will have significant contributions to the values of these functions and their derivatives. Namely, we will consider a high-speed radial oscillation about a point just above the surface with radius $r_{\rm s}=2M$ (in the following, we will always take $r_{\rm s}=1$ for numerical evaluations). We will use three parameters to describe this movement: the distance $d$ of the centre of oscillation to the horizon, the amplitude $A$ and the frequency $\omega$. Then, the radius at which the shell is located will follow the spacetime curve (see fig.~\ref{f2})
\begin{equation}
R(t_-)=r_{\rm s}+d+A\sin(\omega t_-).
\end{equation}
In order to avoid the formation of a horizon and maintain a timelike trajectory, the parameters must satisfy the relations
\begin{equation}
A<d\quad \text{and}\quad A\omega<1.
\end{equation}
We stress once again that the purpose of this study is to gain a better understanding of the relation between dynamical regimes close to horizon formation and the magnitude of semiclassical effects, and not to provide a self-consistent solution with a classical matter content which satisfies some energy conditions. Thus we only impose that the shell be causal, with no further restrictions to its trajectory.\par
Since the trajectory is described in terms of the interior coordinate system, we can obtain the simple expression for the interior velocity parameter
\begin{equation}
\alpha_-=\frac{r_{\rm s}+A\omega\cos(\omega t_-)}{r_{\rm s}-A\omega\cos(\omega t_-)},
\end{equation}
while for $\alpha_+$ we must use eq. \eqref{7}. To evaluate these quantities on the points where the shell trajectory intersects the lines of constant $u$ or $v$ we must solve a transcendental equation, which we will do numerically. First we will obtain the individual values of the functions $g$ and $h$, which represent the change in the coordinate description of outgoing and ingoing radial light rays respectively. Then we will calculate the quotient $g/h$ with $h$ evaluated at a point of entry $v_-$ of a light ray into the Minkowski region and $g$ evaluated at the point of exit $u_-$, which carries information of how light rays suffer a temporal dispersion by passing through this region. Looking at eq. \eqref{2} we can see that an ingoing ray $v_-$ connects with an outgoing ray with $u_-=v_-$, so the quotient we are looking for is $g(v_-)/h(v_-)$. This quantity will also describe the evolution of the $in$ quantum vacuum state, defined at the asymptotically flat region at past null infinity, and its comparison with the $out$ vacuum state, defined at future null infinity.
\begin{figure}
	\centering
	\includegraphics[scale=.5]{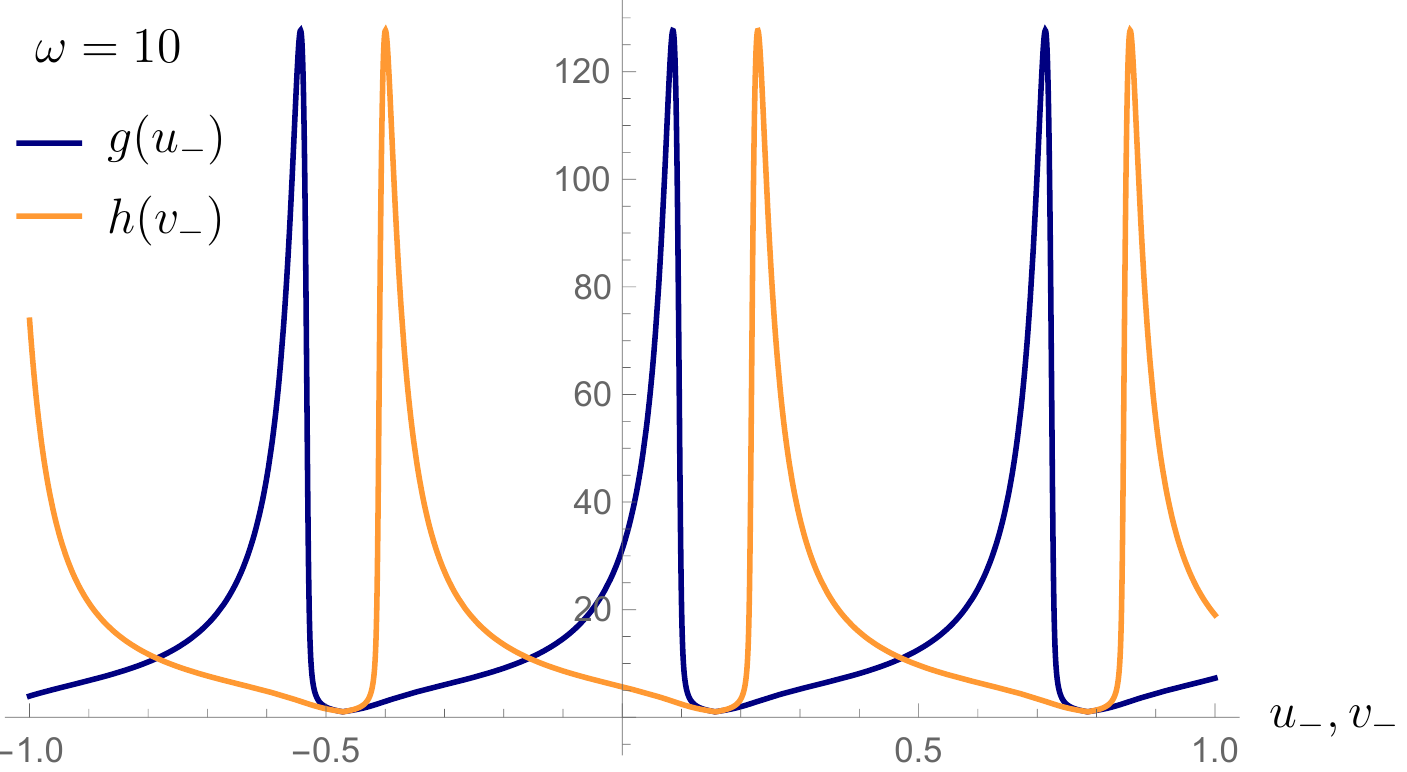}
	\includegraphics[scale=.5]{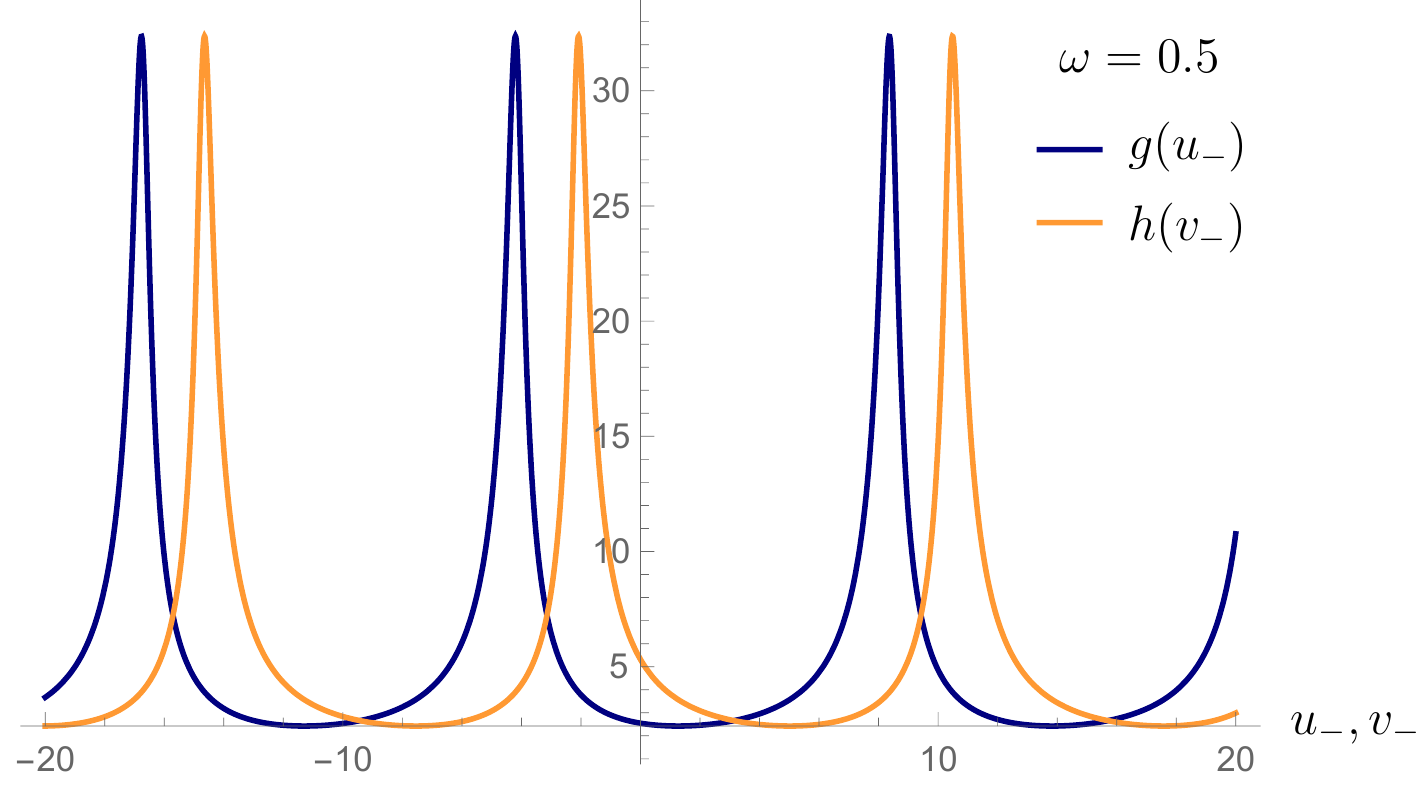}
	\includegraphics[scale=.5]{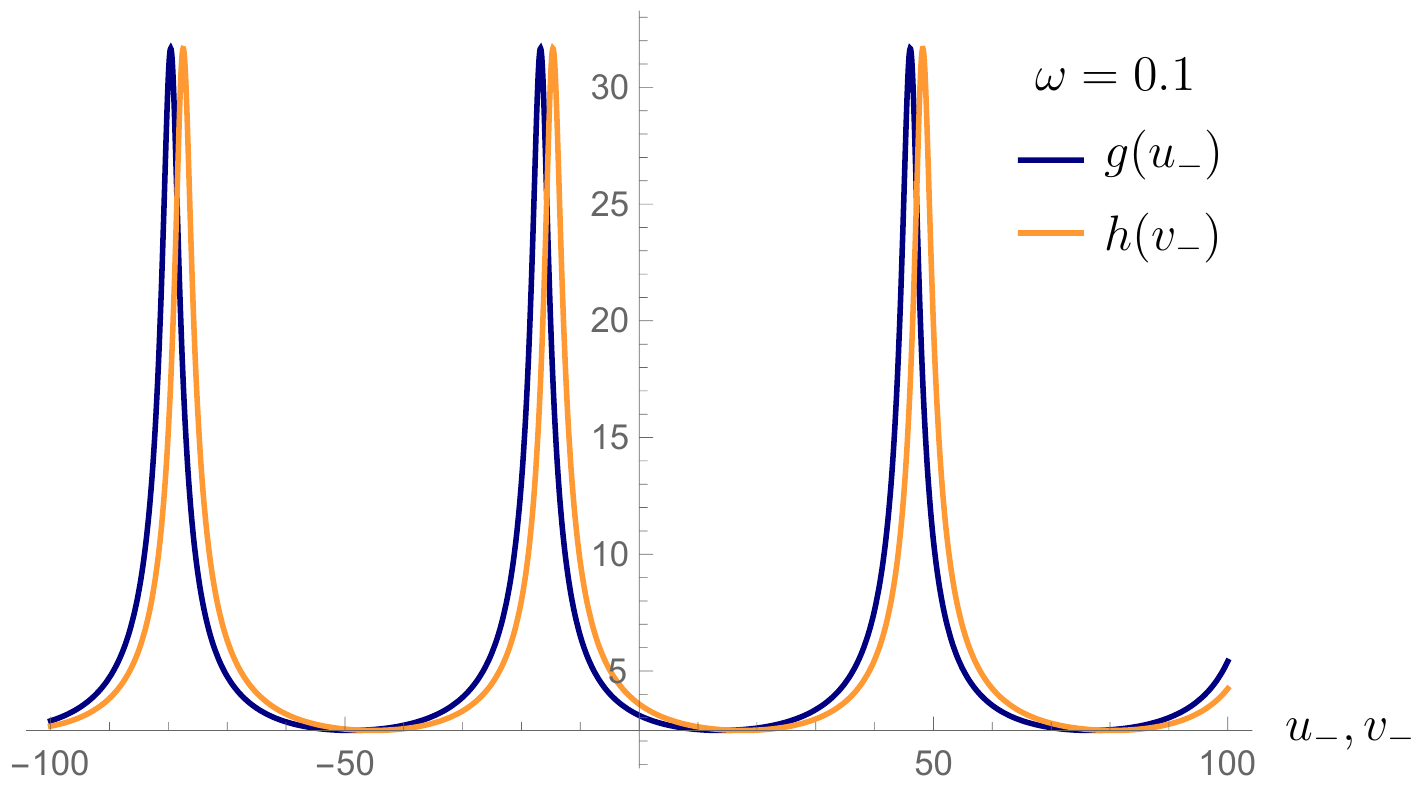}
	\caption{Functions $g$ and $h$ for an oscillation with parameters $d=0.1$, $A=0.099$ and three different frequencies: $\omega=10$, $\omega=0.5$ and $\omega=0.1$. The peaks are produced when the shell is nearly at the closest point to the horizon, as will be discussed below. We observe that at low frequencies the functions practically coincide since the light rays enter and leave the interior region in a time much smaller than $\omega^{-1}$, so the in-crossing and out-crossing dispersion effects would almost cancel out (i.e. $g(u_-)/h(v_-=u_-)\simeq1$). At somewhat larger frequencies the light rays enter and exit at appreciably different points of the oscillation and the functions attain a relative displacement. Finally, at frequencies which make the shell move at nearly light-speed the displacement is greater still, and the peaks become somewhat tilted to one side for each function, due to the fact that the peaks of the sine function in $t_-$ become tilted when seen in the $u_-$ and $v_-$ coordinates (in opposite directions).}
	\label{f3}
\end{figure}
\par
In figure \ref{f3} we observe the values of the functions $g$ and $h$ evaluated at $u_-$ and $v_-=u_-$, representing the dispersion of a light ray when it is exiting and entering the interior region respectively. The net effect, given by $g/h$, reduces to nearly unity when $g(u_-)\simeq h(u_-)$, which occurs when the shell is oscillating very slowly (at low $\omega$) compared to the time it takes for light to cross it (in the static limit, $g/h=1$). At higher frequencies the light rays enter and exit at completely different points of the oscillation, as in the case represented in fig.~\ref{f2}, and the net effect becomes appreciable. It is easy to notice that there are some special cases for this net effect corresponding to different resonances between the oscillation frequency and the crossing time of the light ray: say, when it enters crossing a maximum and also exits crossing one, or crossing a minimum, and a few other such situations. These will be studied in more detail in the following subsection.

\subsection{Resonance between in-crossing and out-crossing effects}
From equations \eqref{6} we can obtain the expression for the total temporal dispersion suffered by a light ray entering the shell at a point ``in" and exiting at a point ``out",
\begin{equation}\label{17}
\frac{du_{+,{\rm out}}}{dv_{+,in}}=\frac{g|_{\rm out}}{h|_{\rm in}}=\frac{\sqrt{f}|_{\rm in}}{\sqrt{f}|_{\rm out}}\left.\sqrt{\frac{\alpha_-}{\alpha_+}}\right|_{\rm out}\left.\sqrt{\frac{\alpha_-}{\alpha_+}}\right|_{\rm in},
\end{equation}
where in the first step we have made use of the fact that for rays reflecting at the origin $dv_-|_{in}/du_-|_{out}=1$, as can be seen from \eqref{2}. We can see again that for a static shell, for which the surface redshift function would be constant and $\alpha_\pm=1$, this quotient reduces to unity. For a moving shell the effects can cancel out again only in one special case, which occurs when not only the $in$ and $out$ redshift functions are the same, but also when $\alpha_-=1$ (and therefore $\alpha_+=1$ as well, as can be seen from eq. \eqref{7}) at both points. For the case of an oscillating shell this can occur only when a light ray exists such that it both enters and exits at a minimum or at a maximum of $R(t_-)$. Then the effects cancel out locally, but they continue being non-trivial for the rest of the light rays. These local resonances are possible only when the frequency, amplitude and distance from the horizon satisfy the relations
\begin{equation}\label{18}
\omega=\frac{n\pi}{r_{\rm s}+d-A}, \quad\text{with }n\text{ integer less than }\quad\frac{r_{\rm s}+d-A}{a\pi},
\end{equation}
for a ray entering and exiting at a minimum, and likewise
\begin{equation}\label{19}
\omega=\frac{n\pi}{r_{\rm s}+d+A}, \quad\text{with }n\text{ integer less than }\quad \frac{r_{\rm s}+d+A}{a\pi},
\end{equation}
for a maximum. These expressions are obtained simply by comparing the ray crossing time and the oscillation periods in the coordinate $t_-$. The upper bound on the values of $n$ comes from the causal restriction $A \omega<1$. For a shell following an arbitrary (known) radial motion such cases can be found just as easily.\par
\begin{figure}
	\centering
	\includegraphics[scale=.59]{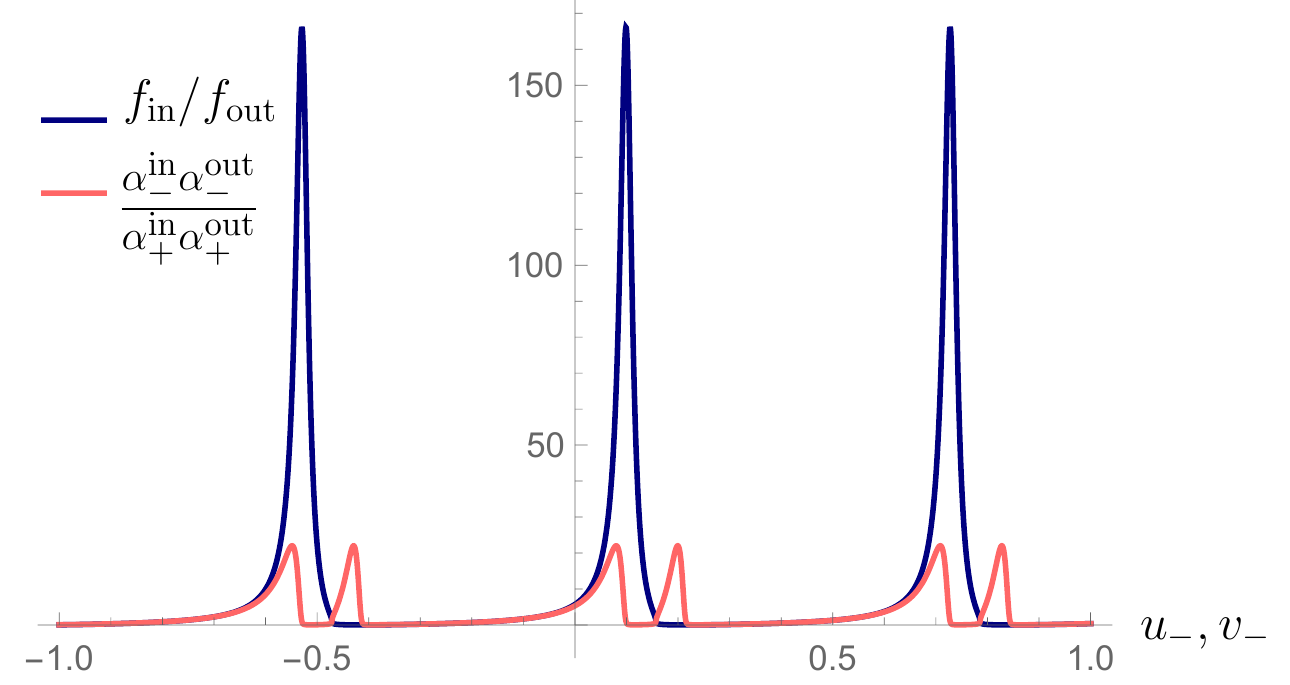}
	\includegraphics[scale=.59]{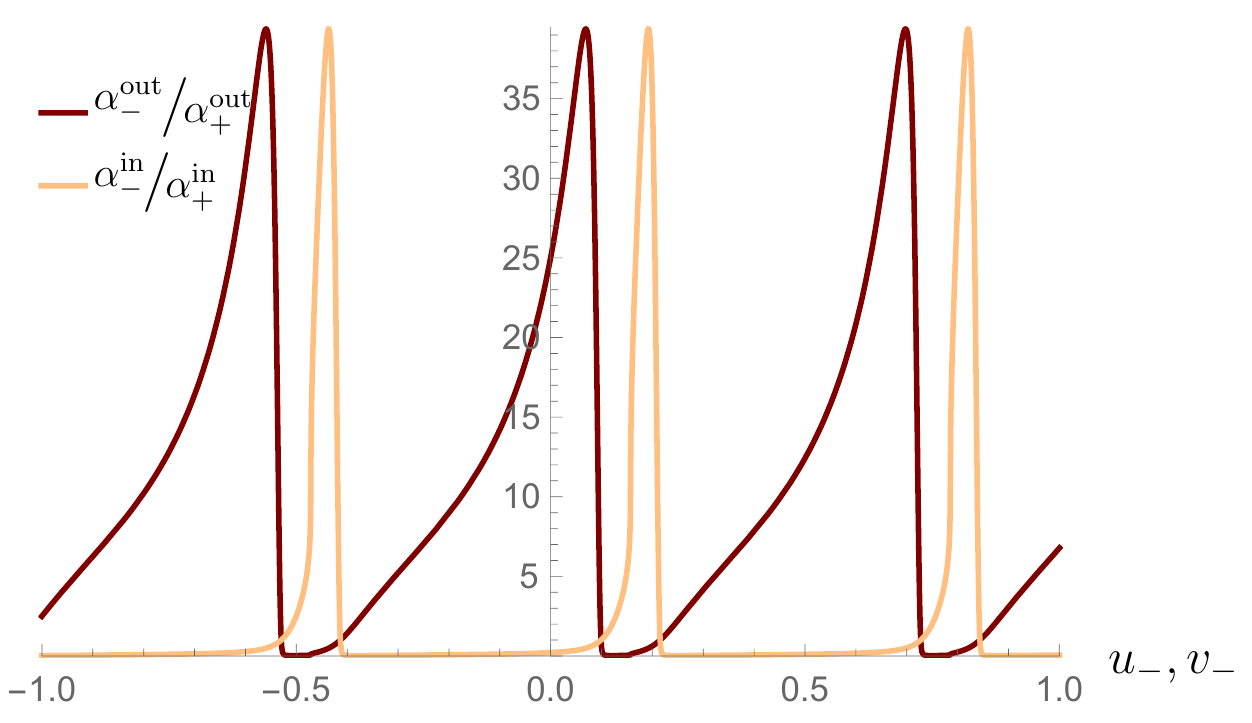}
	\caption{Left: total redshift and Doppler terms plotted separately (the square root of their product gives $du_{\rm out}/dv_{\rm in}$) for on oscillation with parameters $d=0.1$, $A=0.099$ and $\omega\simeq9.996$, given by \eqref{20} with $n=3$. We see that even though the shell reaches 99\% of the speed of light (as seen in the Minkowski coordinates) and the Doppler terms become quite large, the redshift term clearly gives the dominant contribution around its maxima. Right: in-crossing and out-crossing Doppler terms plotted separately. We observe that the peaks and valleys are completely out of phase between the two.}
	\label{f4}
\end{figure}

On the other hand, if we want to see when a maximisation of $du_{\rm out}/dv_{\rm in}$ in eq. \eqref{17} takes place, a more detailed analysis is necessary. First, we may notice that when a light ray enters at a maximum of $R(t_-)$ and exits at a minimum, the total redshift effect is maximised. For such a ray to exist, the relation between the parameters must be
\begin{equation}\label{20}
\omega=\frac{\pi}{2}\frac{2n+1}{r_{\rm s}+d}, \quad\text{with }n\text{ integer less than }\quad \frac{r_{\rm s}+d}{A\pi}-\frac{1}{2}.
\end{equation}
In fig.~\ref{f4} we observe the three terms of the rhs of eq. \eqref{17} plotted (without the square roots) for this case. The peaks of the redshift term, which correspond to precisely the light rays described, reach their highest possible values for the parameters $A$ and $d$ used.
\par
\begin{figure}
	\centering
	\includegraphics[scale=.45]{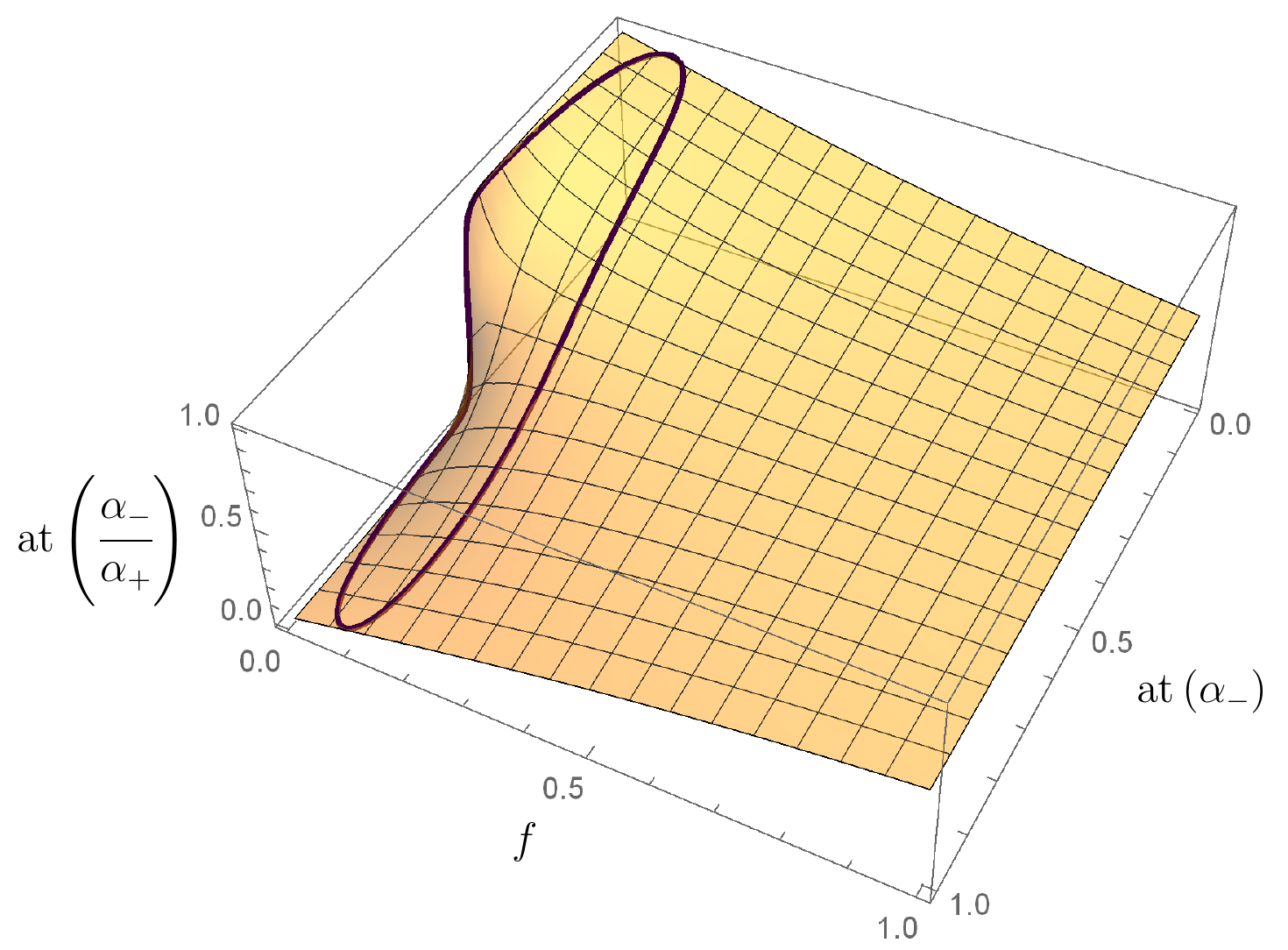}
	\includegraphics[scale=.45]{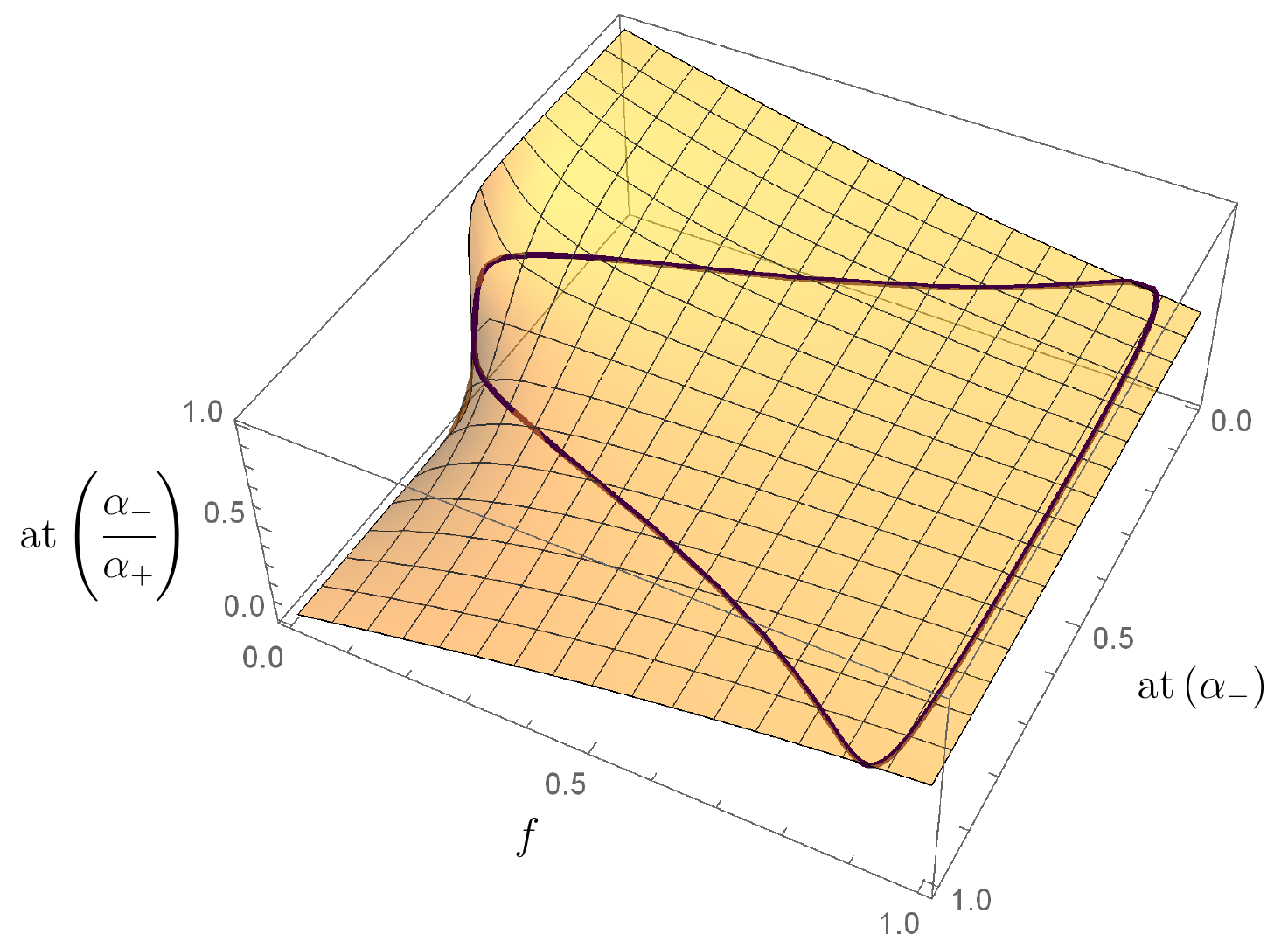}
	\caption{Values of the Doppler term $\alpha_-/\alpha_+$ as a function of $f$ and $\alpha_-$. The axes of $\alpha_-$ and $\alpha_-/\alpha_+$ have been rescaled with a function $\text{at}(x)=\frac{2}{\pi}\tan[-1](x)$ to scale down their whole range into $(0,1)$. The curve drawn on top of the surface on the left represents the values taken during a period of oscillation with parameters $d=0.1$, $A=0.099$ and $\omega=10$, and the curve on the right during an oscillation with parameters $d=10$, $A=9.99$ and $\omega=0.1$. The region with a sharp gradient close to the horizon ($f=0$) is produced around $\alpha_-=1$ ($1/2$ in the graphic), corresponding to the transition from falling inward (during which time $\alpha_-<1$) to going outward (during which $\alpha_->1$). At $f\to1$ the value of the Doppler term tends to 1 smoothly, as the interior and exterior geometries become the same.}
	\label{f5}
\end{figure}

Also in fig.~\ref{f4}, we observe that the individual Doppler terms have distinct maxima. For the in-crossing term the maximum is produced for a ray which enters slightly after the one which maximises redshift (which enters at a maximum of $R$), during the in-fall of the shell. For the out-crossing term it is produced for a ray which exits slightly before redshift maximising one (which exits at a minimum of $R$), so again during an in-fall of the shell. Guided by this result, we can look for the conditions which maximise the individual Doppler terms, and also see whether there is a frequency for the shell at which the two peaks coincide to make a maximum net effect. In fig.~\ref{f5} we can directly see the values which $\alpha_-/\alpha_+$ takes at different redshifts $f$ and velocity parameters $\alpha_-$.\par
At the very minimum of the oscillation of the shell (the closest point to $r=r_{\rm s}$), $\alpha_-/\alpha_+=1$ and there is no Doppler effect for any redshift $f$. When the shell is moving outward ($\alpha_->1$) but is still close to the horizon ($f\ll1$), from eq. \eqref{7} we get
\begin{equation}
\frac{\alpha_-}{\alpha_+}\simeq\frac{\alpha_-^2}{(1-\alpha_-)^2}f,
\end{equation}
that is, at a constant velocity the Doppler term has a linear dependence on the redshift function, with a slope which grows rapidly as $\alpha_-\to1^+$ and which tends to 1 as $\alpha_-\to\infty$. On the other hand, when the shell is falling in, the function close to the horizon can be expressed as
\begin{equation}
\frac{\alpha_-}{\alpha_+}\simeq\frac{(1-\alpha_-)^2}{f},
\end{equation}
which grows parabolically as $\alpha_-\to0$ (as the in-fall speed increases) and hyperbolically as $f\to0$ (as the formation of the horizon is approached).\par
With the above equations and fig.~\ref{f5} we can see that the point of the shell trajectory where the Doppler effect reaches a maximum appears in the $\alpha_-<1$ region, and that its precise position is influenced by two factors: at a constant $f$ it is maximum at the highest velocity (lowest $\alpha_-$), increasing parabolically as $\alpha_-$ decreases, while at a constant velocity it is maximum at the lowest $f$, with a hyperbolic divergence at $f=0$. If $\alpha_-$ approaches 1 while $f$ approaches 0, that is, if the shell tends to a full stop just before the formation of the horizon, then, when $f$ is sufficiently small, the hyperbolic divergence dominates over the parabolic tendency to zero and the maximum is reached at a point very close to the minimum value of $f$, just before the region of very large gradient observed in fig.~\ref{f5} is entered. If, on the other hand, the shell oscillations are produced far away from the Schwarzschild radius $r_{\rm s}$, the maximum Doppler effect is reached closer to the point of maximum in-fall velocity (minimum $\alpha_-$).\par
\begin{figure}
	\centering
	\includegraphics[scale=.59]{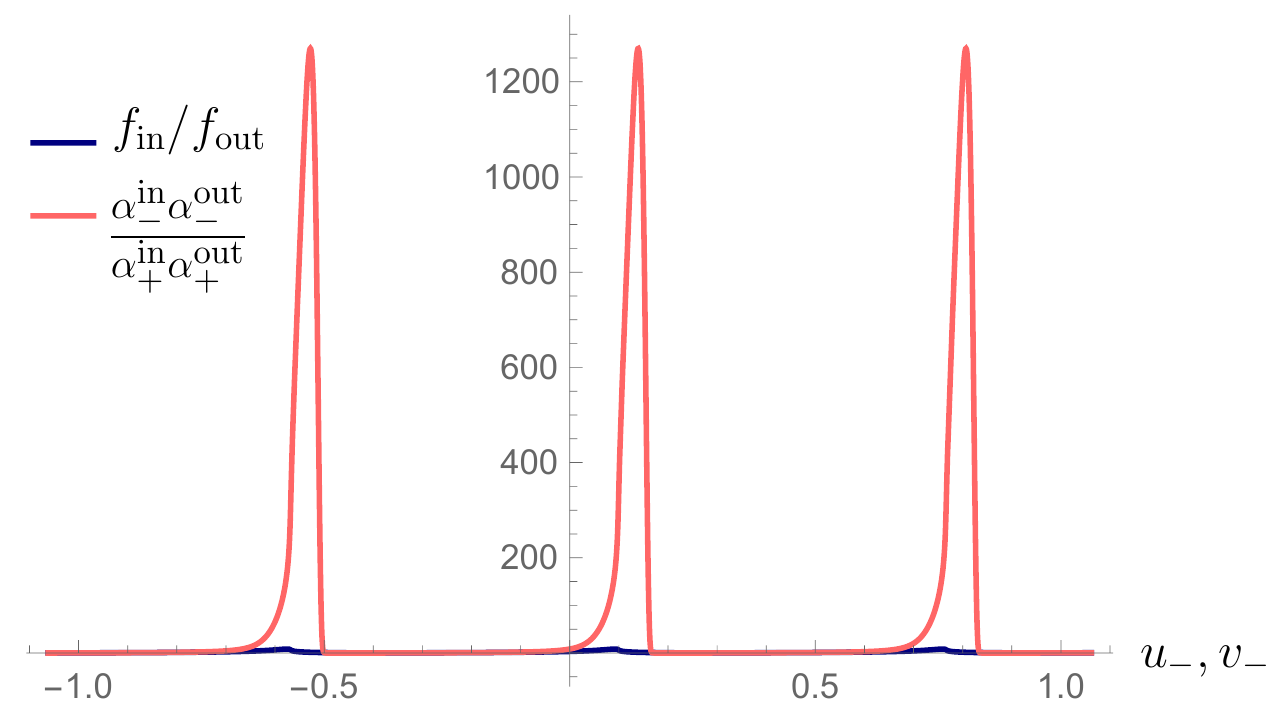}
	\includegraphics[scale=.59]{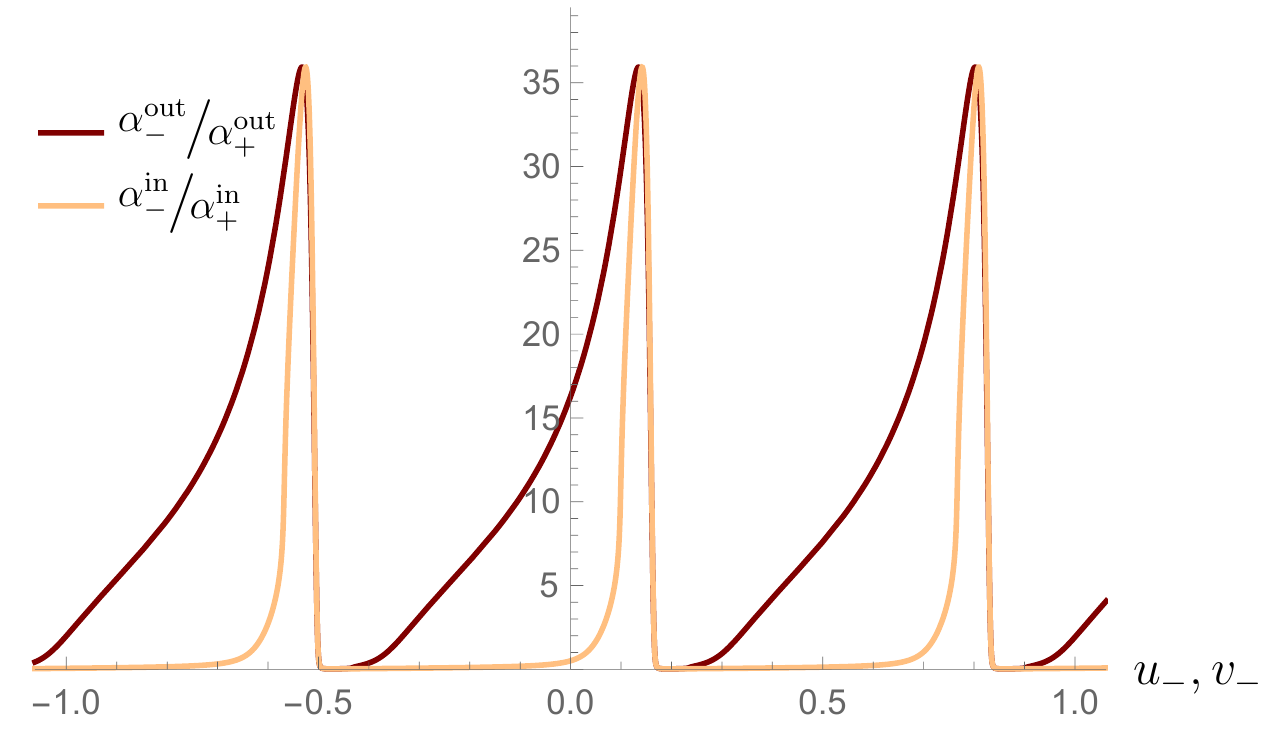}
	\caption{Left: total redshift and Doppler terms plotted separately (the square root of their product gives $du_{\rm out}/dv_{\rm in}$) for an oscillation with parameters $d=0.1$, $A=0.099$ and $\omega\simeq9.42$, given by eq. \eqref{18} with $n=3$. In this case we see the Doppler term clearly dominates. Right: in-crossing and out-crossing Doppler terms plotted separately. We observe the almost-coincidence of the two Doppler peaks, produced for two very close rays passing through the shell slightly before the one which enters and exits at a minimum of the oscillation. This near-coincidence results in the dominance of the net Doppler term in the left graph.}
	\label{f6}
\end{figure}

As an example, in fig.~\ref{f6} we can see the almost-coincidence of the two Doppler peaks (it looks exact in the figure) for an oscillation which bounces at $d-A=10^{-3}r_{\rm s}$, with a frequency $\omega$ which allows rays which enter at a minimum of $R$ to also exit at a minimum. The rays which maximise the in-crossing and out-crossing Doppler effects almost coincide with the ones which cancel out the redshift effect, and even more so with each other. Even when the peaks do not exactly coincide, due to their widths the net Doppler effect given by their product can be very close to its maximum possible value.\par
To conclude, these resonant cases have allowed us to understand the behaviour of the quotient $g/h$ around its highest values, and relate it to specific dynamical regimes of the shell. As we will see, the observed regions of rapid increase or decrease will have significant influence on the behaviour of semiclassical effects.

\subsection{Semiclassical effects}
So far we have studied the dispersion of light rays (or analogously, of modes of the massless scalar field) which cross the oscillating shell and pass through the interior Minkowski region. From these results we can directly calculate the semiclassical quantities discussed earlier, namely the ETF and the RSET. The behaviour of these quantities will be similar to that of the dispersion functions described above, as the former are constructed simply from derivatives of the latter. The structure of peaks and plateaus for each period of the oscillation will merely become more exaggerated for these new functions. For reference, the structure of the ETF due to a single interval of deceleration during collapse has been previously studied with some detail in \cite{Harada2019}. Our study is based on considerably different dynamics, but the results are qualitatively similar.\par
\begin{figure}
	\centering
	\includegraphics[scale=.6]{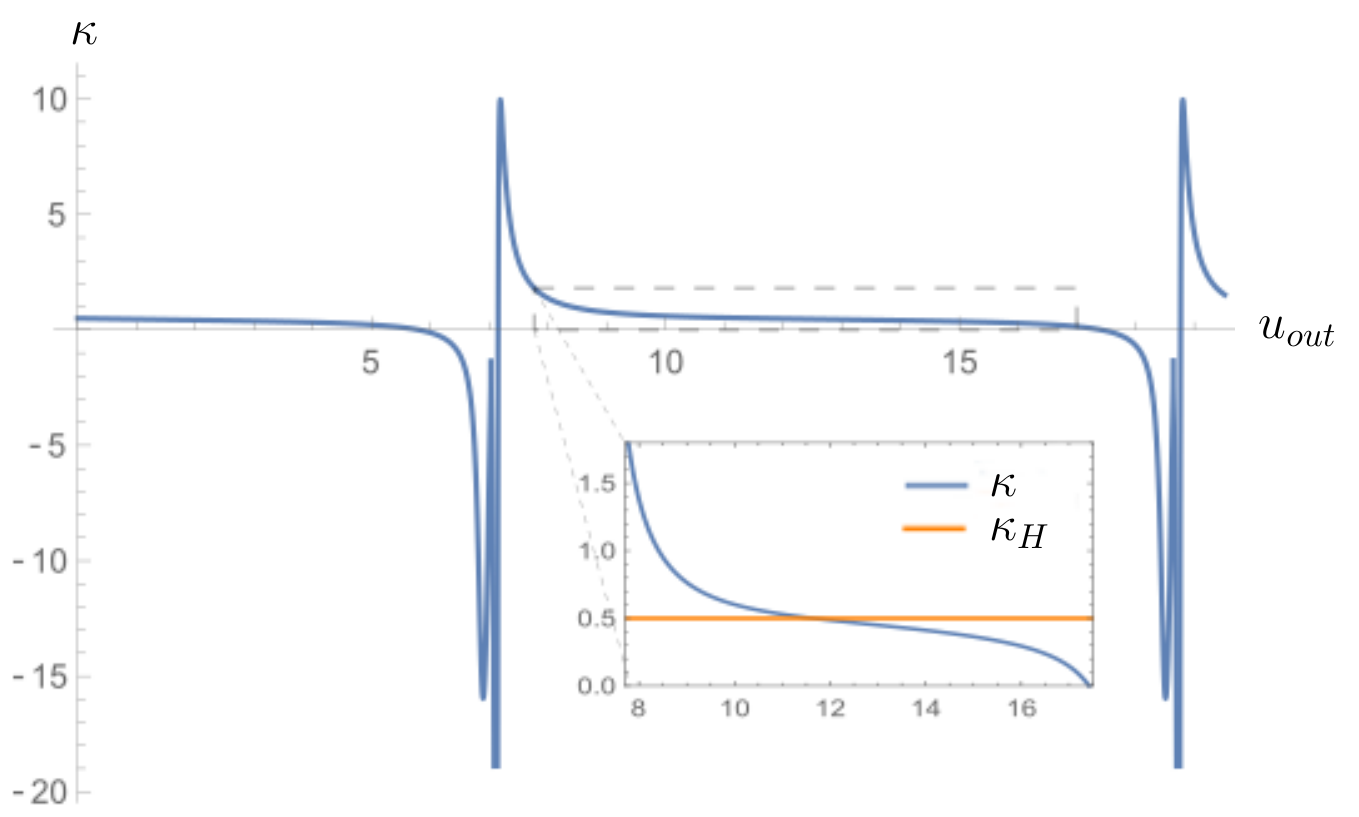}
	\caption{ETF $\kappa_{u_{\rm in}}^{u_{\rm out}}$ produced by an oscillating shell with the same parameters as the ones used for fig.~\ref{f4}, for which the net redshift effect is maximised. The small plot inside the main one is a magnification of the plateau region, along with a comparison with the value $\kappa_{\rm H}$ of the function in the case of Hawking radiation.}
	\label{f7}
\end{figure}

In fig.~\ref{f7} we can see the ETF $\kappa_{u_{\rm in}}^{u_{\rm out}}$, which contains information of the flux of particles seen in the $in$ vacuum state by an inertial observer at future null infinity, calculated with the relation between the $in$ and $out$ coordinates given by the product of the functions plotted in fig.~\ref{f4} through eq. \eqref{17}. As can be guessed by observing the curves in fig.~\ref{f4}, the smaller peaks in $\kappa_{u_{\rm in}}^{u_{\rm out}}$ are produced around the maxima of the Doppler effect contributions. On the other hand, the largest negative and positive peaks are produced on the regions of large gradient on either side of the maximum of the redshift contribution (keep in mind that the horizontal axes of the two plots are rescaled versions of each other). Between each set of peaks there is a region of smoothly decreasing temperature, with values around the Hawking temperature.\par
\begin{figure}
	\centering
	\includegraphics[scale=.6]{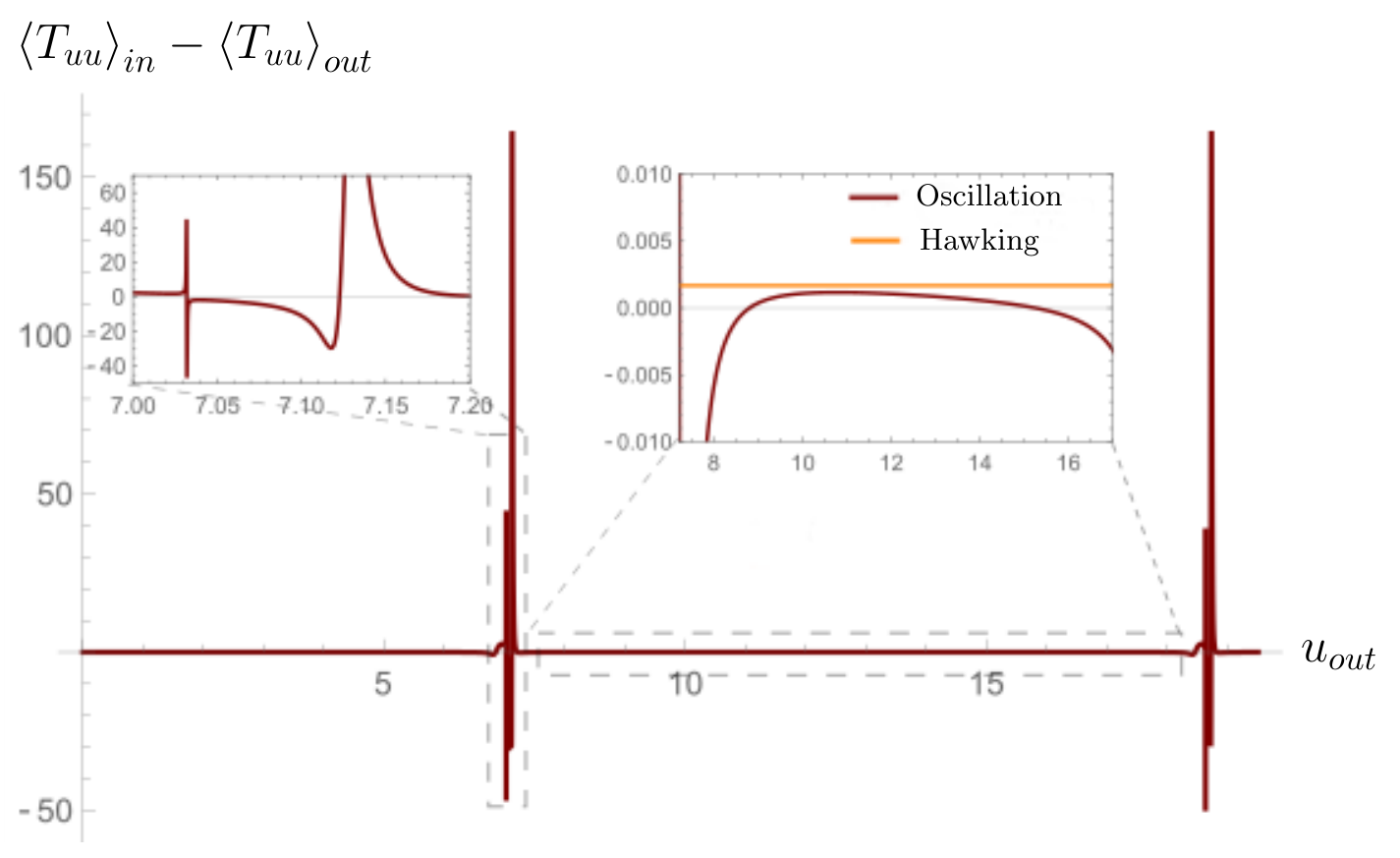}
	\caption{Difference between the $u_{\rm out}u_{\rm out}$ components of the RSET in the $in$ and $out$ vacuum states, corresponding to the outgoing flux of radiation which appears at future null infinity. As in the case of the ETF, we observe periodic peaks, which correspond to the rays which enter at a maximum of the oscillation and exit at a minimum, which maximises the redshift effect, and a more flat intermediate region of values near that of the Hawking radiation flow produced after the formation of a horizon, superimposed in the right zoomed-in rectangle.}
	\label{f8}
\end{figure}

In fig.~\ref{f8} we have plotted the outgoing radiation flux at future null infinity, defined as the difference between $\expval{T_{u_{\rm out}u_{\rm out}}}$ evaluated for the $in$ and $out$ vacuum states. From equations \eqref{27} we see that this quantity depends on $\kappa_{u_{\rm in}}^{u_{\rm out}}$ and its derivative, explaining the somewhat similar, but amplified, characteristics. This quantity alone is representative of the highs and lows of the RSET during the oscillation, since the term which is missing is simply the Boulware vacuum polarisation, which maintains low values in the $u_{out}$ coordinate (outside the horizon it is below the Hawking flux value in fig.~\ref{f8}). It is the $u_{out}$ coordinate itself which tends to become non-regular, leading to a general amplification of both terms (tending to a divergence at the horizon if they do not perfectly compensate each other).\par

\begin{figure}
	\centering
	\includegraphics[scale=.6]{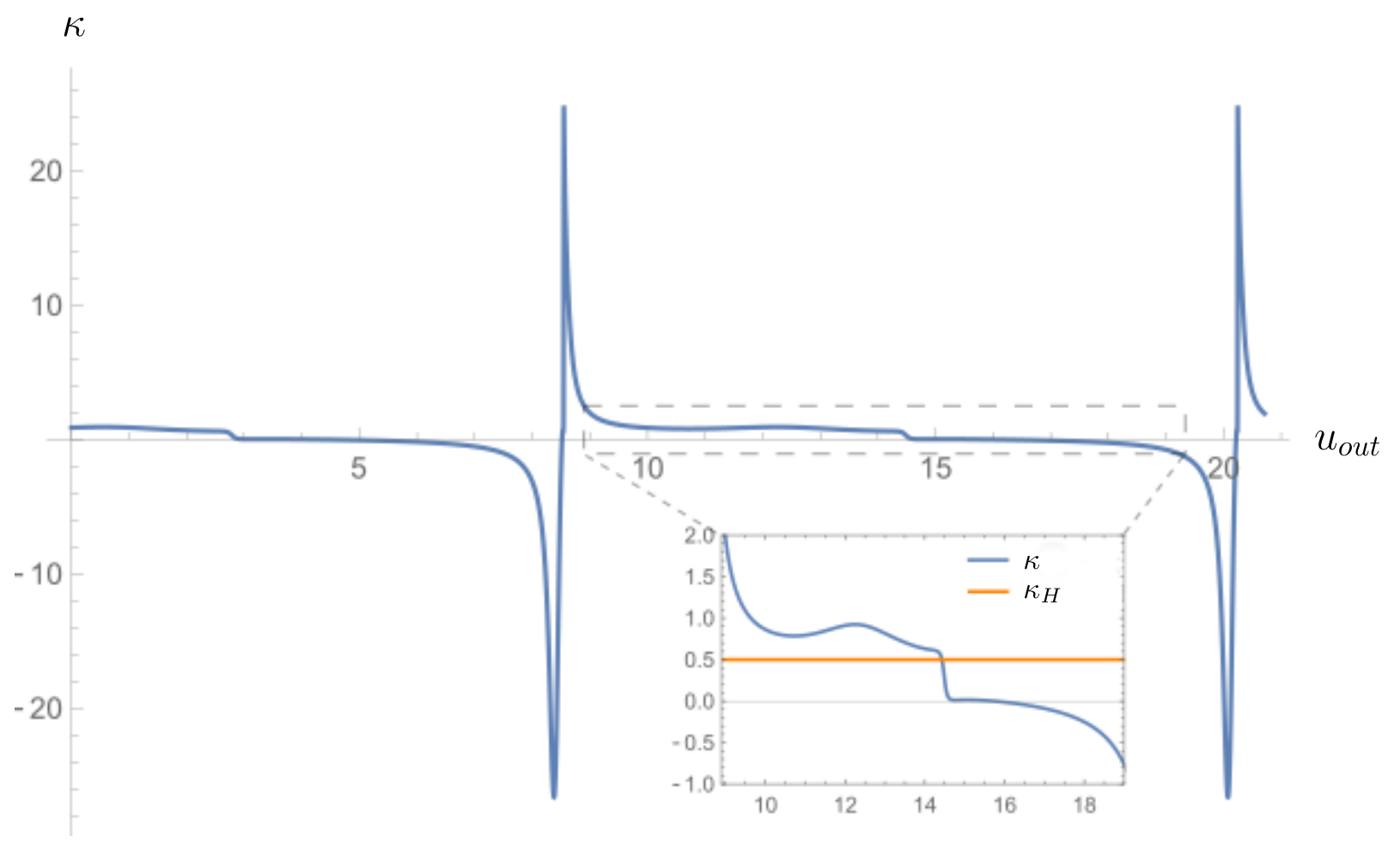}
	\caption{ETF in the outgoing radiation sector, for an oscillation which maximises the net Doppler effect, obtained with the functions plotted in fig.~\ref{f6}.}
	\label{f9}
\end{figure}

In fig.~\ref{f9} we observe the ETF for the oscillation which maximises the net Doppler effect. The two most notable differences with respect to the case which maximises redshift are the somewhat cleaner large peaks, caused by a better overall coincidence in the aspects of the in-crossing and out-crossing effects around the minima of the oscillation, and a less clean intermediate region, caused in turn by a worse coincidence there.\par
\begin{figure}
	\centering
	\includegraphics[scale=0.6]{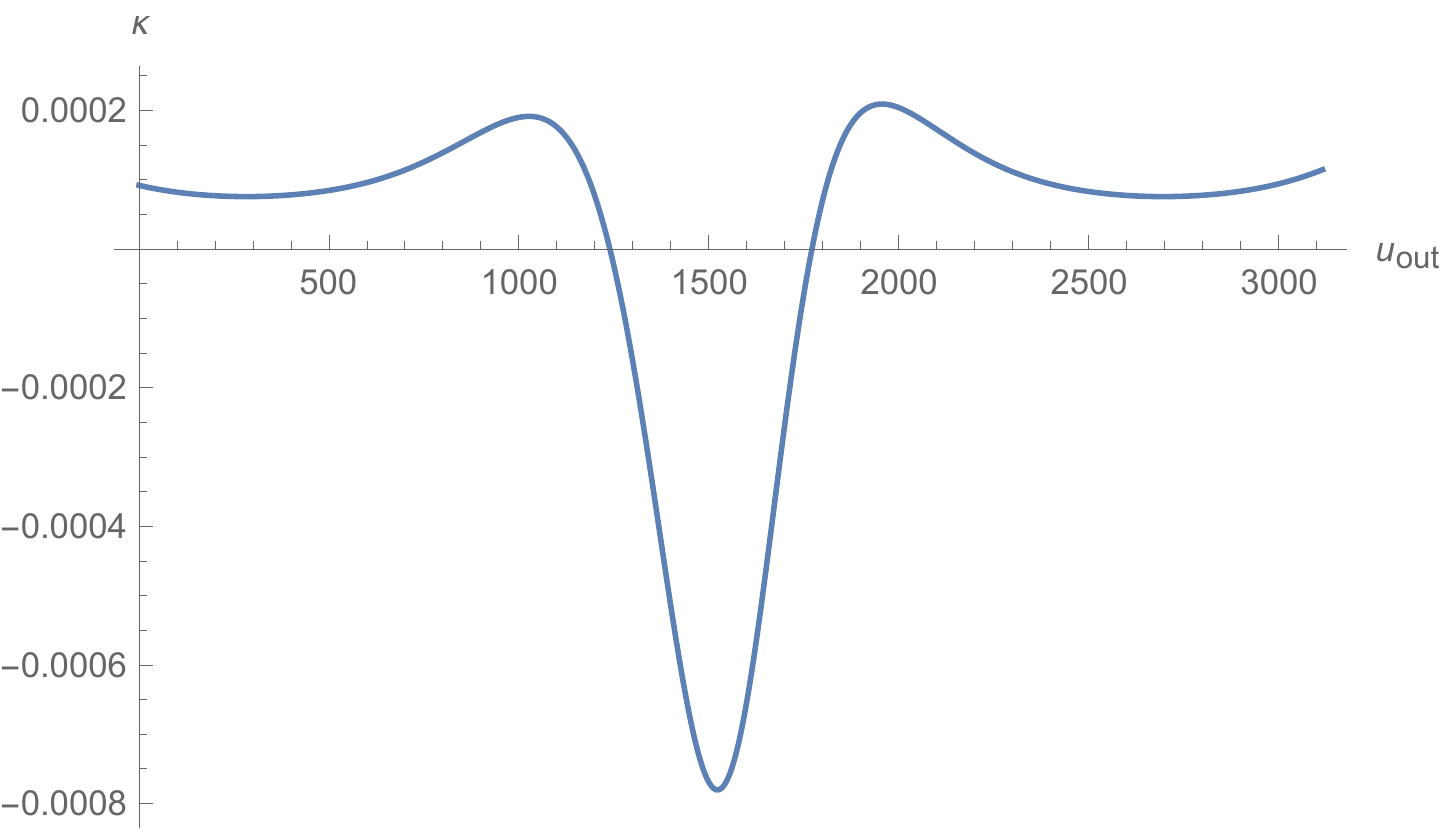}
	\caption{ETF in the outgoing radiation sector, for an oscillation at a low velocity (less then or equal to $0.15\%$ the speed of light) and with a radial proximity to the horizon between $10^{-3}r_{\rm s}$ and $0.2r_{\rm s}$ (the same as in all the cases seen so far). Compared to the cases with higher velocities, we observe a significant decrease of its values and a smoothing of its derivative.}
	\label{f10}
\end{figure}

In order to give a more general picture of the semiclassical effects produced by this type of shell trajectory, we can study the consequence of changing the order of magnitude of each of the oscillation parameters. First, in fig.~\ref{f10} we see the behaviour of the ETF for an oscillation with the same proximity to the horizon (between 0.001 and 0.201 times $r_{\rm s}$) but with a much lower velocity, reaching at most about $0.15\%$ of the speed of light. In this case all semiclassical fluxes are greatly diminished, approaching the static shell limit in which the radiation temperature and flow become zero.\par

Another possibility is to maintain the maximum proximity to the horizon ($10^{-3}r_{\rm s}$) and the large maximum speed ($\sim99\%$ the speed of light), but to vary the amplitude of the oscillation. Decreasing the amplitude leads to a qualitatively similar result for the ETF: each period contains a cluster of large peaks (larger as the amplitude decreases) surrounded by a region of values close to the Hawking temperature. On the other hand, increasing the amplitude to above $0.1r_{\rm s}$ leads to a general decrease in the values at both the peaks and the intermediate regions.\par
The last parameter we can vary is the proximity to the horizon. Understandably, if the shell oscillates very far from the horizon, the ETF and its first derivative become very small, even if the maximum velocities are large. On the other hand, if the shell is close to the horizon, around $10^{-3}r_{\rm s}$ or closer at the minimum of the oscillation, and its amplitude is not very large, then the closer it is, the larger the peaks become, but the intermediate region again remains at values around the Hawking temperature on average.\par

\section{Approaching the horizon asymptotically}\label{s5}
For the case of an oscillating shell studied so far, we have noted that the highest values of the functions which measure the dynamical semiclassical effects are produced when the horizon is approached at very low velocities (around the minima of the oscillation). In fig.~\ref{f5} this was seen through the large gradient in the Doppler term around $\alpha_-=1$ at low values of the redshift function $f$, since both the ETF and the RSET have a dependence precisely on the derivatives of this term. To explore the gap between the cases in which the shell bounces back before forming a horizon, and the ones in which it continues to fall and forms a black hole (which will be the subject of the next section), we can study shell trajectories which tend to the $r=r_{\rm s}$ surface asymptotically, i.e. those that approach the $f\to0$ and $\alpha_-\to1^-$ limit monotonously and reach it in an infinite regular time parameter. In \cite{BLSV04} it was first shown that configurations of this sort can lead to Hawking-like radiation with arbitrarily long duration without necessarily forming any type of horizon. Then in \cite{BLSV06} the same authors analysed more detailed configurations having in mind an analogue gravity setting. The present study reproduces those results and extends them further, with a more generalised approach in the construction of the geometries.\par
\begin{figure}
	\centering
	\includegraphics[scale=.7]{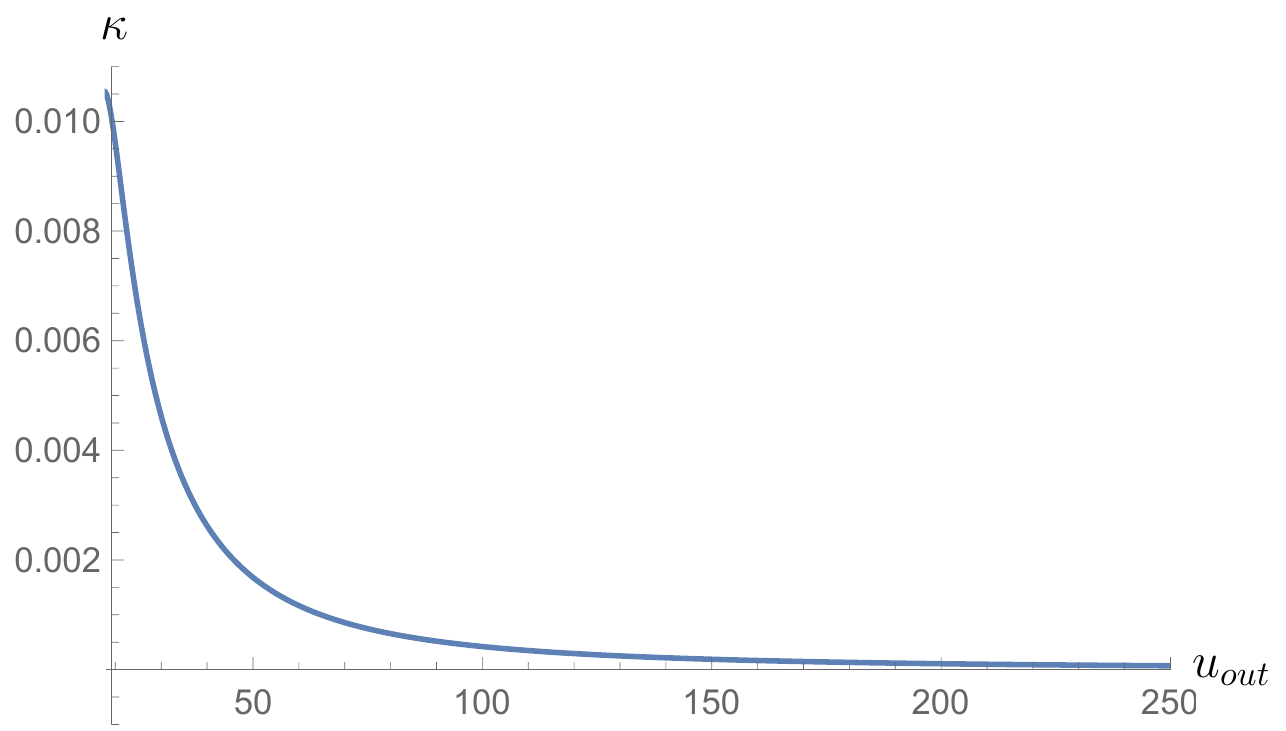}
	\caption{ETF for the outgoing radiation sector in the case of a thin spherical shell which follows a collapse trajectory $R=1+e^{-v_-}$ (as always, in $r_{\rm s}=1$ units). We observe that the function tends to zero, indicating that there is no finite outgoing radiation flux asymptotically, meaning the dynamical $in$ vacuum tends to the Boulware vacuum above the shell surface.}
	\label{f11}
\end{figure}

To start with, we can use the same formalism as in the previous section, approximating a spherical distribution of matter with an infinitesimally thin shell. Following the same scheme as before, we can decide on a shell trajectory and then see how the ETF behaves, from which we can also guess how the RSET is modified in the dynamical $in$ vacuum with respect to the Boulware vacuum. Taking, for example, the trajectory $R=r_{\rm s}(1+e^{-v_-/r_{\rm s}})$ we obtain the ETF plotted in fig.~\ref{f11}, which rapidly tends to zero.\par
When the ETF tends to zero, it leaves the RSET in the exterior Schwarzschild region to approach the divergence it has in the Boulware vacuum as the shell approaches $r_{\rm s}$, as can be seen clearly from eqs. \eqref{27}. In the Boulware vacuum, when matter has crossed its Schwarzschild radius there is a $1/f(r-r_{\rm s})$ type of divergence (with $f$ being the redshift function) \cite{Boulware,Visser96}. In this case, as the ETF between the $in$ and Boulware vacua tends to zero, the values of the RSET for the $in$ vacuum in the exterior $r>R$ region approach those for the Boulware vacuum. The plot in fig.~\ref{f11} therefore represents the decreasing difference in time between the physical RSET and a quantity which at the shell surface increases as $1/f(R)\sim 1/(R-r_{\rm s})\sim e^{v/r_{\rm s}}$, i.e. exponentially in time.\par
In fact, it turns out that for any such asymptotic approach of the shell to its Schwarzschild radius the result is qualitatively the same: an ETF which tends to zero and an RSET with rapidly increasing values. To understand why this occurs, we can look at the definition of the ETF in eq. \eqref{26}, and see that it only has a finite (constant) asymptotic value if the relation $du_{\rm in}/du_{\rm out}$ is asymptotically an exponential in $u_{\rm out}$. Additionally, if the shell has positive mass and is in continual in-fall, then light rays become more dispersed in time after travelling through it and escaping. This implies that the asymptotic relation between the coordinates must be of the type $du_{\rm in}/du_{\rm out}\sim e^{-ku_{\rm out}}$, with $k>0$. Integrating this relation, we see that $u_{\rm out}$ reaches infinity for a finite value of $u_{\rm in}$, i.e. the outgoing light rays need to be trapped inside a finite spatial region after some moment. The dynamics of an infinitesimally thin shell cannot trap light rays in such a way without also forming a horizon in finite time, so a non-zero asymptotic flux of field particles is only obtained when the collapsing shell forms a proper black hole. In that case the relation between the $in$ and $out$ coordinates becomes
\begin{equation}
\frac{du_{\rm in}}{du_{\rm out}}\to (const.)\,e^{-u_{\rm out}/2r_{\rm s}}
\end{equation}
at large $u_{\rm out}$'s, which gives the Hawking temperature result $\kappa_{u_{\rm in}}^{u_{\rm out}}=1/2r_{\rm s}$ previously mentioned. If the Schwarzschild radius is only reached asymptotically, no light ray ever gets trapped.\par
So from this result it may seem that if a distribution of matter approaches the formation of a horizon only asymptotically, then the exterior vacuum polarisation would always tend to its values in the Boulware vacuum. However, we cannot generalise the result obtained for an infinitesimally thin shell in such a way. Very often, shells provide a good and simple model for collapse scenarios, giving quite similar results to more realistic models of matter, for instance making any shell a thick one. But when studying asymptotic results in the vicinity of a horizon, the effect of even the tiniest width for the shell can change the result entirely, as we will see below. This can be seen as an interesting illustration of the idea that horizons can act as a magnifying glass of high energy physics (in this case represented by the detailed structure inside a thin shell).

\subsection{Light-ray trapping without horizon formation}
\begin{figure}
	\centering
	\includegraphics[scale=.9]{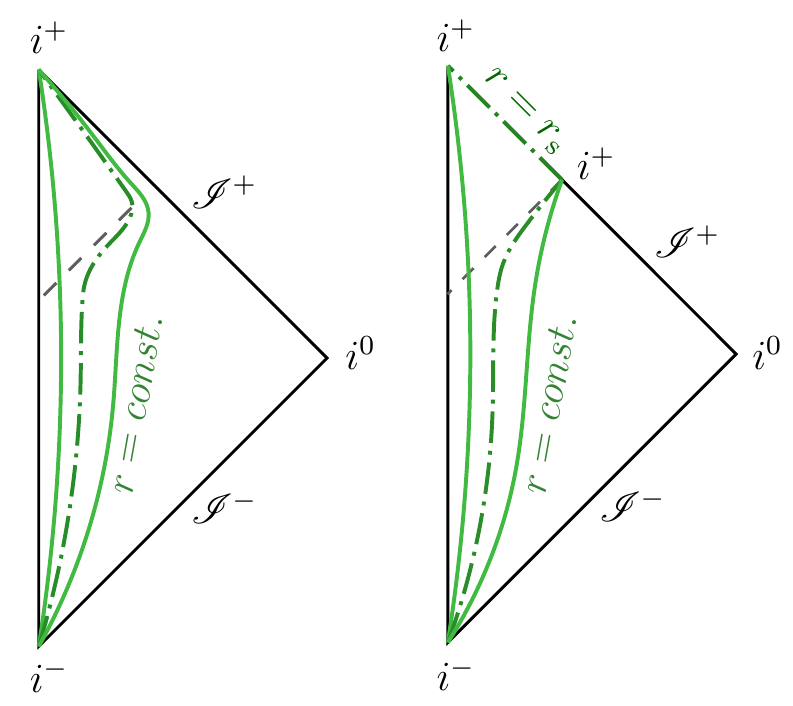}
	\caption{Conformal diagrams of two spacetimes in which light rays are trapped inside a finite spatial region. The curves indicate surfaces of constant radius. The dash-dotted curve is the Schwarzschild radius $r=r_{\rm s}$. The diagram on the left represents confinement which only lasts a finite time, without the formation of an event horizon. The diagram on the right represents confinement which lasts all the way to the asymptotic future null region, forming an event horizon. These two cases have an initial region of identical semiclassical (and classical) behaviour. In the second case an inner Cauchy horizon may also form, resulting in an extendibility of the geometry analogous to that of an extremal charged black hole.}
	\label{f11-1}
\end{figure}

Essentially, the reason why the result is different when a finite-volume matter-filled region is introduced (instead of a thin shell) is because the resulting geometry \emph{can} trap light rays inside a finite spatial region without forming an apparent horizon in finite time, only tending to its formation asymptotically in time. More specifically, in such a case outgoing light-ray trajectories would remain confined inside the Schwarzschild radius for an arbitrarily long, or even infinite period of time (as measured by the regular coordinate $v_{in}$), tending to escape only asymptotically in time and thus never doing so. Fig.~\ref{f11-1} shows a conformal diagram of this type of spacetime, for both the case in which light rays eventually escape as well as the case in which they do not. In terms of the expansion of these null geodesics, this situation would be characterised by an expansion which tends to zero at the Schwarzschild radius $r=r_{\rm s}$ only asymptotically in time, as opposed to the standard black-hole formation scenario, in which it becomes zero after a finite time. This confined state of the light rays at least gives the ETF the possibility of having a finite asymptotic value. Whether that cuts off the growing values of the RSET as the Boulware divergence is approached is another matter still.\par
To clarify, if this light-trapping behaviour were to be maintained asymptotically, even if no apparent horizon were formed in finite time, the null surface described by the first trapped radial ray would in fact become an event horizon. However, by manipulating the geometry further one could stop this asymptotic tendency at any time and let the trapped light rays out of their spatial confinement. The interesting thing is that before one does so, a Hawking-like flux of radiation can be maintained for an arbitrarily long period of time, without the need to form any sort of horizon. On the other hand, if this (quasi)thermal flux were slightly different from the case of Hawking radiation (or absent altogether, as in the thin shell case), the RSET would again tend to a divergence at the Schwarzschild radius.\par
To see when each of these possible outcomes actually takes place, we will present and categorise a large family of spherically-symmetric geometries which trap light rays while having only an asymptotic tendency to form a horizon. For the complete picture, consider a spacetime given by an exterior patch of a Schwarzschild geometry and an interior patch with an arbitrary (spherically-symmetric) distribution of matter, given by the line element in advanced Eddington-Finkelstein coordinates
\begin{equation}\label{29}
ds^2=-f(v,r)dv^2+2y(v,r)dvdr+r^2d\Omega^2,
\end{equation}
where $f$ and $y$ are arbitrary functions which depend on the characteristics of the matter content. The two regions are separated by an in-moving spherical surface located at a radius $R(v)$. For convenience we will define the coordinate $d\equiv r-r_{\rm s}$, which is just the radial distance from the Schwarzschild radius. Outgoing radial light rays in the interior part of the geometry will follow the trajectories given by the differential equation
\begin{equation}\label{30}
d'(v)=\frac{1}{2}\frac{f(v,d+r_{\rm s})}{y(v,d+r_{\rm s})},
\end{equation}
where $'$ denotes the derivative with respect to $v$. For the exterior geometry $y(v,r)=1$ and $f(v,r)=f(r)=1-r_{\rm s}/r$. We will define the generalised redshift function in both regions as
\begin{equation}
F(v,r)\equiv\begin{cases}
1-\frac{r_{\rm s}}{r},&r> R(v),\\
\frac{f(v,r)}{y(v,r)},&r\le R(v).
\end{cases}
\end{equation}
In the absence of an apparent horizon, $F$ will be positive everywhere. At the interface $r=R$ we will assume that it is at least continuous, and that there its value is a minimum of the function in the radial direction, which tends to zero asymptotically in the temporal direction. With this setup, it turns out that whether or not light rays get trapped in the interior region depends only on $F$ at $R$ and its first non-zero spatial derivative on the interior side. Thus, it is completely independent of the exterior geometry, so we can afford to be a bit lax with the matching conditions and only require continuity of the metric for now.\par
We can expand the generalised redshift function $F$ in a power series in the coordinate $d$ around the curve $d_R(v)\equiv R(v)-r_{\rm s}$ (approaching from the inside, and assuming analyticity there) and write eq. \eqref{30} as
\begin{equation}\label{32}
d'(v)=\frac{1}{2}\frac{d_R(v)}{r_{\rm s}+d_R(v)}+k_1\left[d_R(v)-d(v)\right]+k_2\left[d_R(v)-d(v)\right]^2+\cdots,
\end{equation}
where the first term ensures continuity with the metric in the exterior Schwarzschild region. The coefficients $k_i$ can, in principle, also be variable in time, but we will focus on cases in which the first non-vanishing one remains constant (or sufficiently close) at large times. Since we will only focus on asymptotic solutions, its possible early-time variability and the values of the higher order coefficients will not be relevant.\par
We will define three categories for the possible functions $d_R(v)$, covering all monotonous asymptotic approximations to the $d=0$ surface (the Schwarzschild radius). Then we will see the most general conditions the coefficients $k_i$ must satisfy in each case for light rays to get trapped.

\begin{figure}
	\centering
	\includegraphics[scale=.6]{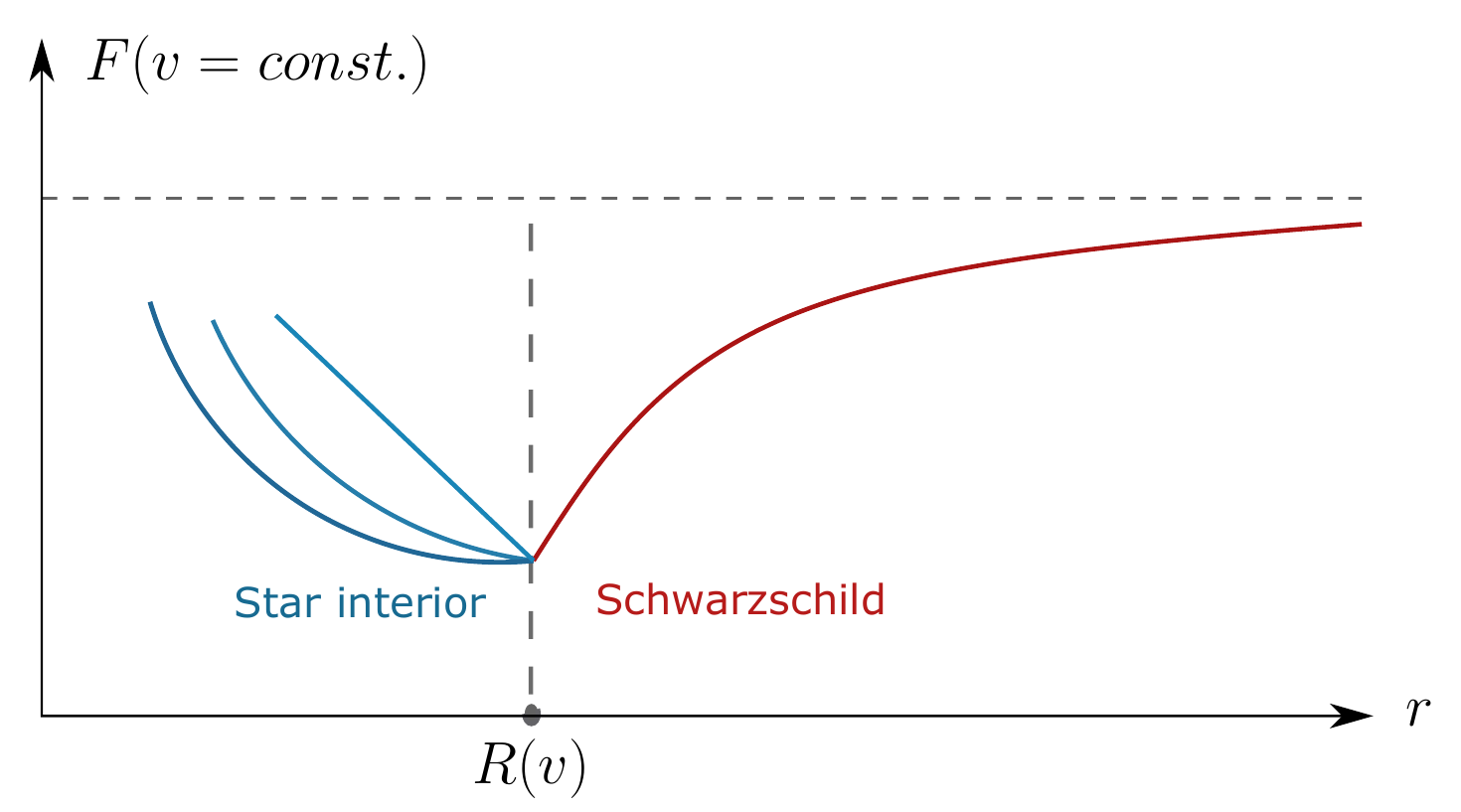}
	\caption{Qualitative plot of different possible redshift functions $F$ at radii around the surface $R(v)$ at some instant $v=const.$ during the collapse.}
	\label{f12}
\end{figure}

\paragraph{Sub-exponential approach:}
The first type of surface trajectory we consider is an approach to the Schwarzschild radius with a distance which decreases as the inverse of a polynomial,
\begin{equation}
d_R(v)=r_{\rm s}\left(\frac{r_{\rm s}}{v}\right)^n,\qquad \text{with }n\text{ real and positive}.
\end{equation}
Let us call $m$ the degree of the first non-zero coefficient of the series expansion \eqref{32}, i.e.
\begin{equation}\label{34}
d'(v)=\frac{1}{2}\frac{1}{1+(v/r_{\rm s})^n}+k_m\left[r_{\rm s}\left(\frac{r_{\rm s}}{v}\right)^n-d(v)\right]^m+\cdots.
\end{equation}
We have said that the redshift function $F$ has a minimum at $d_R$, so $k_m>0$. The value of $m$ can be thought of as a measure of the width of this minimum (on the inside), as the larger it is, the smoother the function becomes around $d_R$. In the limit $m\to\infty$ it becomes constant in $d$ at equal times, making its approach to zero extend to all points of the interior region. Each of these radial points would then mark an asymptotic marginally trapped surface.\par
In order to look for trapped solutions (remaining inside $d<0$ but close to $d_R$ at large times), we have to make some assumption about their asymptotic behaviour. If we assume that the $(-d)^m$ term dominates on the rhs of \eqref{34}, we obtain that if
\begin{equation}\label{35}
m-1>\frac{1}{n-1}
\end{equation}
is satisfied, there are trapped asymptotic solutions of the type
\begin{equation}\label{36}
d\sim -\frac{1}{(k_m(m-1))^{\frac{1}{m-1}}}\frac{1}{(v-c)^{\frac{1}{m-1}}},
\end{equation}
where $c$ is an integration constant. In fact, for these solutions $m$ can also be any real number greater than 1. On the other hand, if we assume that the terms with $1/v^n$ dominate, under the same condition \eqref{35} we obtain another trapped solution
\begin{equation}
d\sim -\frac{1}{2(n-1)}\frac{r_{\rm s}^n}{v^{n-1}},
\end{equation}
which, compared to the previous solutions through the inequality \eqref{35} can be seen to be asymptotically closer to $d=0$, and therefore corresponds to the first trapped light ray. All light rays passing through the distribution of matter in a radial direction after the one which has the above asymptotic solution become trapped inside.
\begin{figure}
	\centering
	\includegraphics[scale=.5]{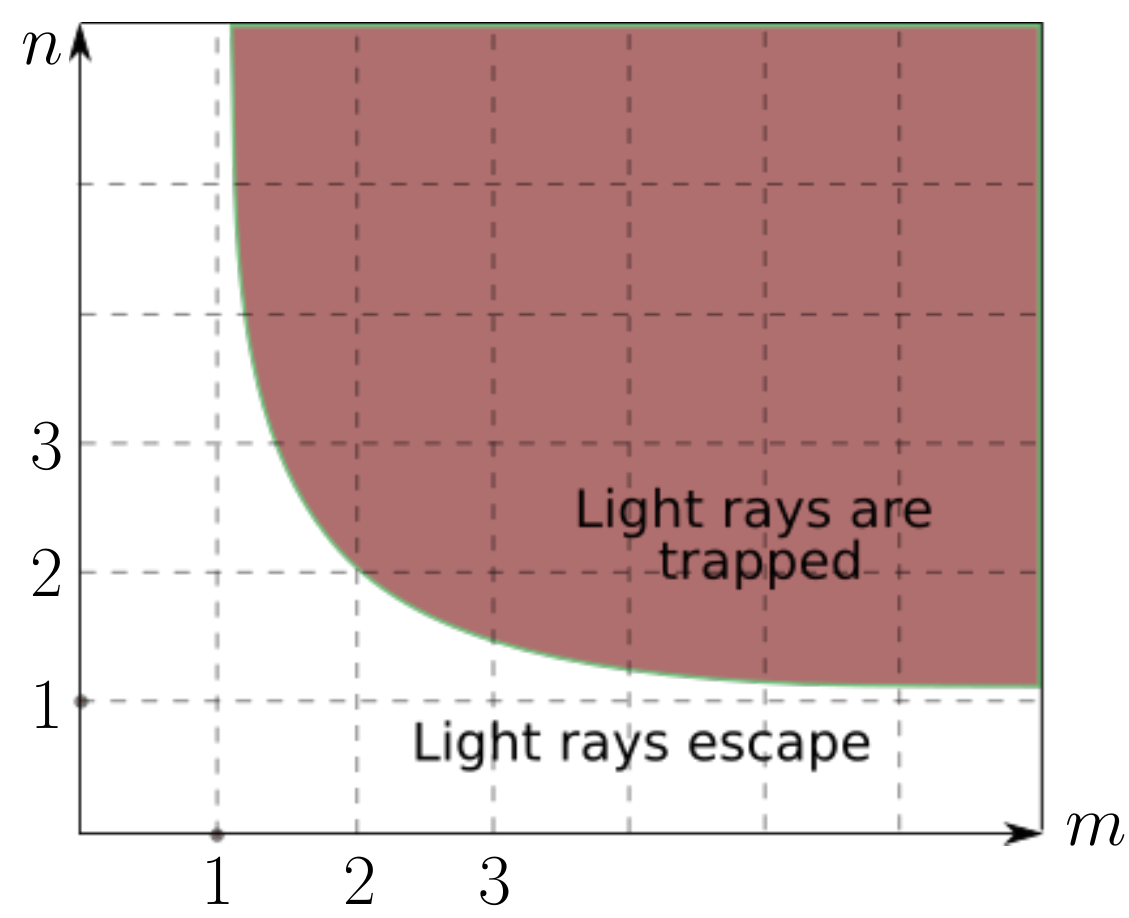}
	\caption{Region of the space of parameters $m, n$ which satisfies the inequality \eqref{35}, allowing for light rays to be trapped inside the asymptotic horizon.}
	\label{f13}
\end{figure}

\paragraph{Exponential approach:}
We now consider a surface trajectory of the type
\begin{equation}\label{38}
d_R(v)=r_{\rm s}e^{-\gamma v},\qquad\text{with }\gamma\text{ real and positive}.
\end{equation}
In this case we have the same differential equation \eqref{34}, only with exponentials instead of polynomials. If the degree $m$ of the first non-zero term in the expansion is strictly greater than 1, then the solution which is asymptotically below $d=0$ but gets closest to it, i.e. the first trapped ray, can be obtained by assuming that the rhs of the equation is dominated by terms of order $e^{-\gamma v}$. The result is
\begin{equation}
d\sim -\frac{r_{\rm s}}{2\gamma}e^{-\gamma v}.
\end{equation}
On the other hand, if $m=1$, then asymptotically we must consider the terms with $e^{-\gamma v}$ and $d(v)$ in the differential equation. Then it turns out that there are asymptotic trapped solutions only if
\begin{equation}
\gamma>k_1.
\end{equation}
Their expressions up to order $e^{-\gamma v}$ are
\begin{equation}\label{41}
d\sim ce^{-k_1v}-\frac{1}{2}\frac{1+2k_1}{\gamma-k_1}e^{-\gamma v},
\end{equation}
where again $c$ is an integration constant. The first trapped light ray corresponds to $c=0$, while subsequent trapped ones correspond to values $c<0$. The limit $c\to 0^+$ corresponds to the last escaping ray.

\paragraph{Super-exponential approach:}
If $d_R(v)$ approaches zero faster than an exponential (e.g. a Gaussian), then light rays are trapped for any $m\ge1$. The first trapped solution is asymptotically proportional to the integral of $d_R(v)$ (the error function for a Gaussian).

\subsection{Asymptotic temperature}
Having seen a quite general family of geometries for which escaping light rays do get trapped, we can now connect them with the outside and check how semiclassical effects behave around the Schwarzschild radius, where an apparent horizon is formed asymptotically. At first, we will allow a discontinuity in the first derivatives of the metric components at the surface $R$ and see what values the ETF takes. After that we will consider a case in which the transition is smoothed out (with zero spatial derivatives for $F$ on both sides of its minimum, making it behave like the redshift function in an extremal charged black-hole formation process) and see how the result changes.\par
\begin{figure}
	\centering
	\includegraphics[scale=.55]{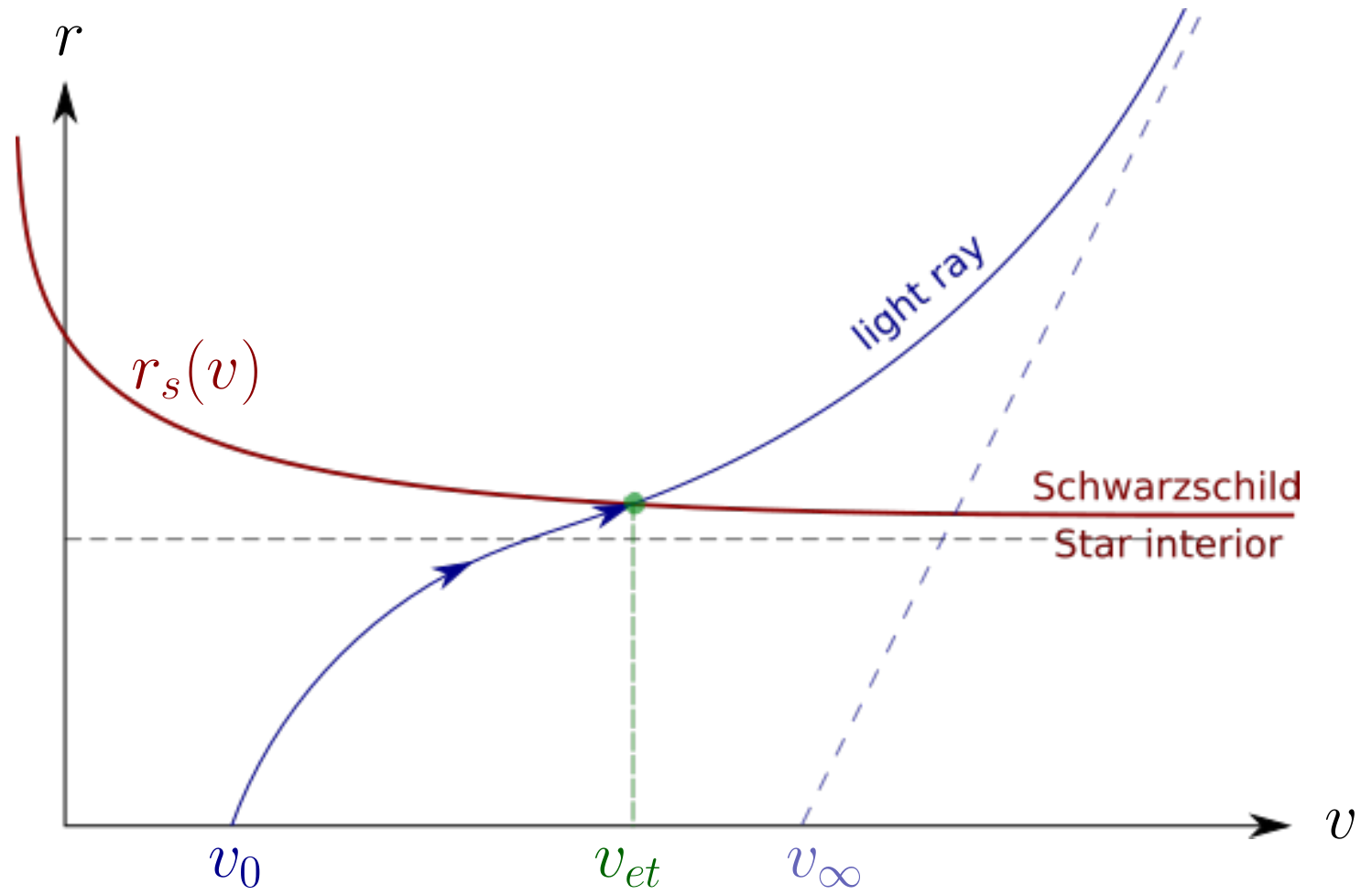}
	\caption{The trajectory of an outgoing light ray which is emitted at $r=0$, escapes the surface of the matter distribution $R$ and reaches infinity, tending to a straight-line trajectory.}
	\label{f14}
\end{figure}

Let us first trace the trajectories of the light rays in the exterior geometry, from the moment in which they cross the surface located at $R$ (at the escape time $v_{\rm et}$) until they reach future null infinity, and explain how the ETF is calculated. To label the specific light rays at future infinity, we will use the parameter $v_\infty$ given by the origin of the asymptotic straight line which the ray tends to follow, as can be seen by looking at fig.~\ref{f14}. Integrating the outgoing null geodesic equation in the Schwarzschild region, we can obtain this parameter as a function of the surface point,
\begin{equation}\label{42}
v_\infty=v_{\rm et}-2R(v_{\rm et})-2r_{\rm s}\log[R(v_{\rm et})-r_{\rm s}].
\end{equation}
When a horizon is formed in finite regular time (and remains present forever), $v_\infty$ diverges as the argument of the logarithm tends to zero, while the rest of the terms $v_{\rm et}-2R(v_{\rm et})$ remain finite and negligible. Then,
\[R(v_{\rm et})-r_{\rm s}\sim e^{-v_\infty/2r_{\rm s}},\]
and the ETF $\kappa_{u_{in}}^{u_{out}}=1/2r_{\rm s}$ is simply the multiplicative constant in this exponential when the other term (in this case $R(v_{\rm et})-r_{\rm s}$) tends to a simple zero. In this case we see that the internal structure of the collapsing matter distribution is not in any way reflected in the asymptotic ETF, hence why the Hawking temperature of black holes depends only on their Schwarzschild radius.\par

On the other hand, when a horizon is formed asymptotically, both the logarithm and $v_{\rm et}$ diverge on the rhs of \eqref{42}. If they both diverge logarithmically, then the ETF is modified and can now reflect some of the characteristics of the interior geometric structure. If either of them diverges hyperbolically, then the ETF shuts down altogether, tending to zero roughly as $1/v_\infty$.\par
This last behaviour is precisely what occurs in the case for the sub-exponential approach of the surface to the Schwarzschild radius, in which $v_{\rm et}$ in \eqref{42} diverges hyperbolically. When the ETF tends to zero, the RSET on the surface approaches its Boulware divergence, as was the case for the infinitesimal shell.\par
For the case of the exponential approach, the result is the first of the above-mentioned: both terms in \eqref{42} diverge logarithmically. For $k_1\neq0$ (and $\gamma>k_1$, as is necessary for light rays to be trapped), the contribution of each divergence (in the same order as in \eqref{42}) is
\begin{equation}\label{43}
v_\infty\sim-\frac{1}{\gamma-k_1}\log(\epsilon)-\frac{2\gamma r_{\rm s}}{\gamma-k_1}\log(\epsilon),\qquad\epsilon\to 0,
\end{equation}
where this last limit means an approach of the arguments of both logarithms to a simple zero. The asymptotic ETF is then
\begin{equation}
\kappa_{u_{\rm in}}^{u_{\rm out}}=\frac{\gamma-k_1}{2\gamma r_{\rm s}+1},
\end{equation}
which is always less than the Hawking temperature, approaching it only in the $\gamma\to\infty$ limit (an infinitely quick collapse). If $k_1=0$, the result is the same as the above taking $k_1\to 0$, also making the lower bound on the possible values of $\gamma$ zero.\par
Lastly, for the case of the super-exponential approach to the Schwarzschild radius, the dominant divergence in \eqref{42} is a logarithmic one with the same coefficient as in the case of horizon formation in finite time, leading to an asymptotic Hawking temperature with the same value. As an example, let us consider a surface trajectory of the type
\begin{equation}
d_R=r_{\rm s}e^{-(\gamma v)^n},\qquad \text{with }n\ge 2.
\end{equation}
Then the diverging terms in $v_\infty$ corresponding to $v_{\rm et}$ and the logarithm are respectively
\begin{equation}
\frac{1}{\gamma}\left[\log(\epsilon)\right]^{1/n}-2r_{\rm s}\log(\epsilon),\qquad \epsilon\to 0,
\end{equation}
so the dominant divergence is simply the same logarithm as in the Hawking case.\par
We can therefore say that if a collapse is sufficiently quick, then, as far as long-term semiclassical effects above the horizon are concerned, there is no difference from the case of the formation of a black hole in finite time: the Boulware divergence is canceled out and there is a flux of Hawking radiation. On the other hand, if the collapse is slower, then the asymptotic ETF decreases according to the speed of collapse and to one particular characteristic of the internal structure: the first spatial derivative of $F$ at the surface. We can say that anything further in this region remains invisible to the ETF, just as the whole structure was invisible for a quicker collapse. For a sufficiently slow collapse, when the ETF becomes zero, the internal structure again becomes hidden asymptotically, only this time the semiclassical effects become indistinguishable not from the case of a dynamic black-hole formation, but from the case of a static black hole.

\subsection{A smooth transition: the extremal black hole}
Up to this point we have considered a generalised redshift function $F$ with a minimum which has a discontinuity in the first derivative, as is seen in fig.~\ref{f12}. The slope on the outside has always been finite, given by the derivative of the Schwarzschild redshift function, and has appeared implicitly in the calculations through the quantity $r_{\rm s}$. For the slope on the inside we have analysed the cases in which it may be finite or zero.\par
\begin{figure}
	\centering
	\includegraphics[scale=.7]{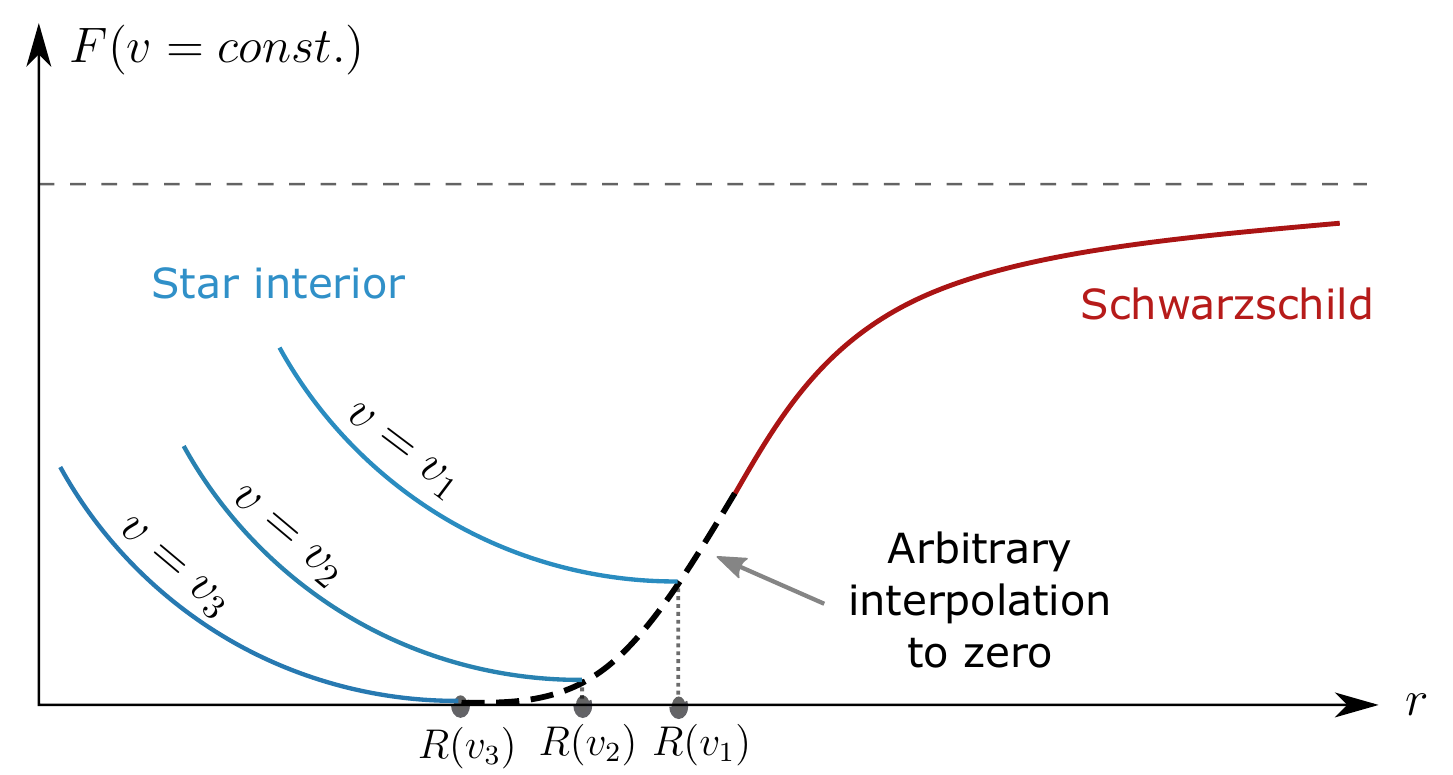}
	\caption{Generalised redshift function $F(v,r)$ with a minimum at the interface between the interior and exterior geometry, located at the point $R(v)$. For the interior geometry three sections of constant time are represented, $v_1<v_2<v_3$. The section at $v_3$ represents the behaviour at times close to infinity. The exterior geometry is static. When $R(v)$ approaches $r_{\rm s}$, the function is continuous but has an otherwise arbitrary behaviour at both sides of the minimum. Far away from $r_{\rm s}$ it transitions into the Schwarzschild redshift function.}
	\label{f15}
\end{figure}

We will now explore the case in which both the slope on the inside and on the outside of the minimum may be zero. Particularly, we will modify the external static geometry being revealed beyond the surface $R(v)$ in such a way that the slope (and subsequent derivatives) on the outside can have an arbitrary value at horizon formation, maintaining however an asymptotically Schwarzschild structure, as shown in fig.~\ref{f15}. We are going to show that, in accordance with the results in \cite{BLSV06}, the asymptotic ETF will be zero if the slope on the outside is zero, even for an exponential approach of $R$ to the horizon.\par

To show this, let us trace the trajectories of light rays in this new exterior geometry, from the moment in which they cross $R$ (at the escape time $v_{\rm et}$). For the region close to $r_{\rm s}$, we can write the differential equation which governs their movement as a power series in their distance $d(v)$ from this point (in the same way we expanded it in the distance $d_R(v)$ in \eqref{32} for the interior geometry),
\begin{equation}
d'(v)=\frac{1}{2}F(v,r_{\rm s}+d(v))=k_1d(v)+k_2d(v)^2+\cdots.
\end{equation}
If the exterior geometry is static, then the coefficients $k_i$ are constant, although for our purposes we will only need the first non-zero one to be asymptotically constant, similarly to our previous calculations for the interior region. Let us call this first non-zero coefficient $k_m$, so that the light rays close to $d=0$ will move according to
\begin{equation}
d'(v)\simeq k_md(v)^m.
\end{equation}
For a Schwarzschild exterior $k_m=k_1=1/(2r_{\rm s})$. In general, if $m=1$ then the solutions of the above equation are
\begin{equation}
d(v)\simeq d_R(v_{\rm et})e^{k_1(v-v_{\rm et})}
\end{equation}
for rays crossing the surface at $d_R(v_{\rm et})=R(v_{\rm et})-r_{\rm s}$. On the other hand, for $m>1$ the solutions are
\begin{equation}
d(v)\simeq\frac{1}{\left[d_R(v_{\rm et})^{-(m-1)}-k_m(m-1)(v-v_{\rm et})\right]^{\frac{1}{m-1}}}.
\end{equation}\par
When the exterior geometry was Schwarzschild, we calculated light-ray dispersion through the variation in the quantity $v_\infty$, defined from the integration of null trajectories through the whole exterior region, up to infinity. We did so because this parameter offered the most obvious relation with the label $u_{out}$ which is used to define the $out$ vacuum state (they are in fact proportional to each other). However, to calculate the ETF we only need to study the divergent part of the dispersion, which occurs long before the light rays reach infinity. In fact, for the behaviour of the ETF at large times we only need to trace their trajectories up to an arbitrarily small distance $\varepsilon$ away from the surface where the horizon forms asymptotically, i.e. up to $r_{\rm s}+\varepsilon$.\par
\begin{figure}
	\centering
	\includegraphics[scale=.5]{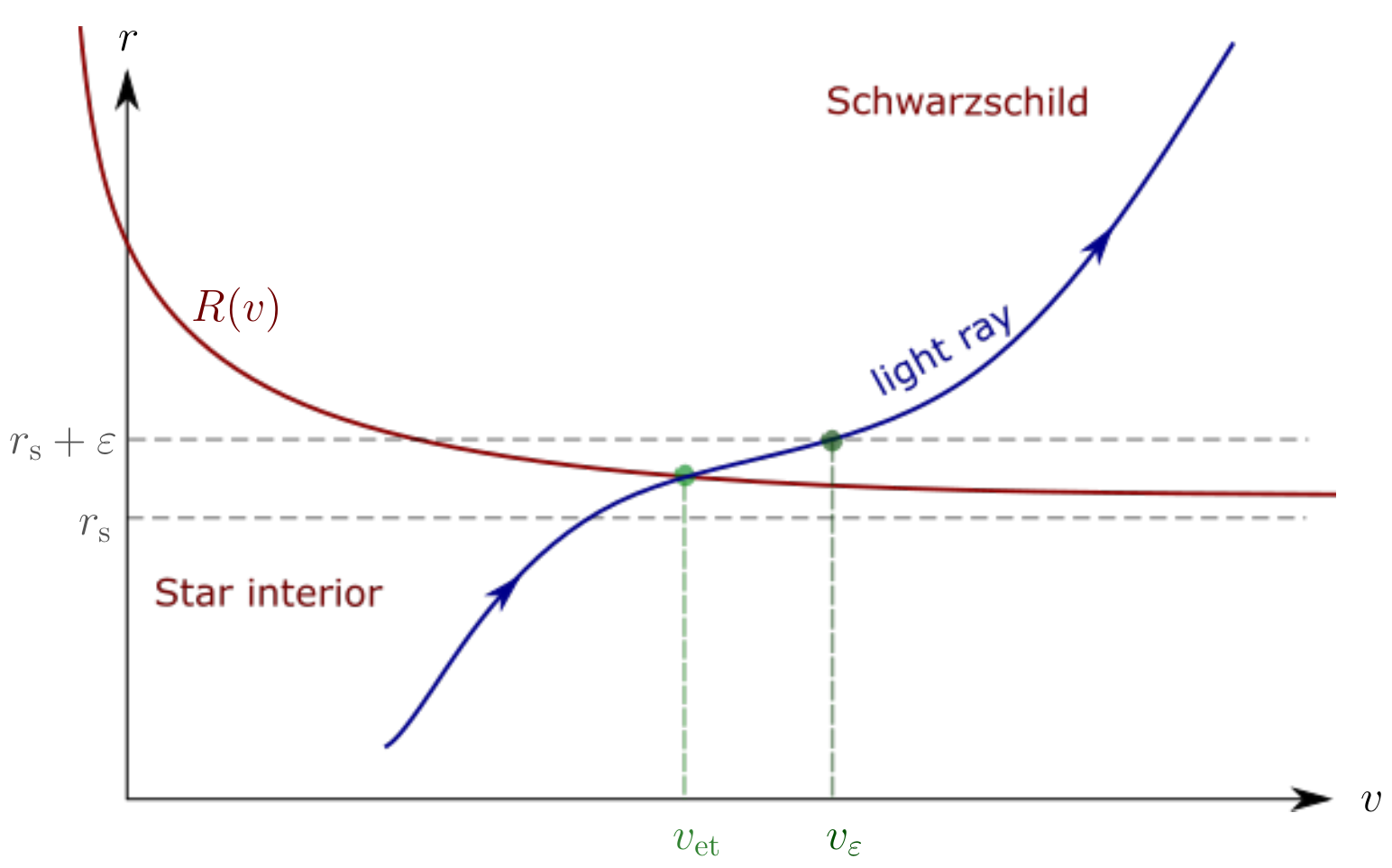}
	\caption{The path of a light ray which escapes from the surface $R$ at time $v_{\rm et}$ and crosses $r_{\rm s}+\varepsilon$ at time $v_\varepsilon$.}
	\label{f16}
\end{figure}

We will define a new parameter $v_\varepsilon$ as the moment light rays cross the $r_{\rm s}+\varepsilon$ surface (which is always outside the surface $R(v)$ at large enough times), as shown in fig.~\ref{f16}. This parameter will take the place of $v_\infty$ in the study of the divergence in the dispersion of the trajectories of light rays which get arbitrarily close to the first trapped one. For the above solutions with $m=1$ we have
\begin{equation}
v_\varepsilon\simeq v_{\rm et}+\frac{1}{k_1}\log(\frac{\varepsilon}{d_R(v_{\rm et})})\sim v_{\rm et}-\frac{1}{k_1}\log[d_R(v_{\rm et})],
\end{equation}
where the second relation shows that at large times, so long as $\varepsilon$ is finite, the value of $v_\varepsilon$ is in fact independent from this distance parameter. Approaching the last escaping light rays we have $d_R(v_{\rm et})\to 0$ and $v_{\rm et}\to \infty$. From the results in the previous subsection, we know that if the divergence in $v_{\rm et}$ is logarithmic or quicker, then the ETF has a finite asymptotic value. On the other hand, if $v_{\rm et}$ diverges more slowly, then the ETF shuts down.\par

From the solutions with $m>1$ we obtain
\begin{equation}
v_\varepsilon\sim v_{\rm et}+\frac{1}{k_m(m-1)d_R(v_{\rm et})^{m-1}}.
\end{equation}
In this case, no matter how quickly $v_{\rm et}$ diverges, the parameter $v_\varepsilon$ always has a dominant hyperbolic divergence, making the ETF shut down asymptotically in time as $1/v_\varepsilon$.\par
This result shows that if this slope of $F$ on the outside is asymptotically zero, then the ETF tends to zero even for an arbitrarily quick collapse. Consequently, the RSET in the $in$ vacuum tends to its values in the Boulware vacuum, approaching a divergence as $F$ tends to zero. This case should not be confused with the one of the thin shell approaching the Schwarzschild radius asymptotically, as although the ETF and RSET have the same asymptotic behaviour, light rays behave quite differently (in one case they get trapped and in the other they do not).\par
A more similar case in which semiclassical effects have the same long-term behaviour is that of a spacetime in which an extremal charged black hole forms in finite time. In it, the generalised redshift function has a smooth minimum (with zero slope on both sides) which, unlike in our model, reaches zero in finite time (and stays at zero from then on). In this case there is again an absence of an asymptotic flux of particles at future null infinity \cite{Hiscock1977,Liberati2000}.

\section{Horizon formation at different velocities}\label{s6}
Up to this point we have analysed the consequences of staying in the region of large gradient in fig.~\ref{f5} outside the Schwarzschild radius, but approaching it asymptotically. We have also considered a variation of this problem involving a more general distribution of matter. The final step in our study is to see exactly what happens when the Schwarzschild radius \emph{is} crossed in finite null time. Particularly, we have already shown for this case that the asymptotic ETF is always $1/2r_{\rm s}$, and it is a well-known fact that this provides an additional term in the RSET with respect to its Boulware vacuum value that precisely regularises the divergence at the horizon \cite{DFU}.\par
In this section we will be interested in semiclassical effects produced in a finite time interval around the formation of a horizon by a shell collapsing \textit{at different velocities lower than the speed of light} at the moment of crossing the horizon. Past studies in this direction, although detailed, have usually involved only a shell collapsing at light-speed (e.g. \cite{SC2014}), justified by the fact that during astrophysical black-hole formation, the velocity of falling matter is expected to be high when crossing the horizon. By contrast, as mentioned earlier, the goal of this work is to thoroughly study the semiclassical effects produced in more general dynamical situations.\par
Since in this case the asymptotic solutions for both the ETF and RSET are known, we will be more interested in short-term dynamical effects. At horizon formation, large values of the RSET are to be expected if the $in$ vacuum approximates the static Boulware vacuum in some way (say, in the case of a very slow collapse). Therefore, it is at the horizon itself where we might expect the most clear estimate of how large semiclassical effects can become. We will thus be interested in obtaining the total values of the RSET components there. To give them a more physical interpretation, we will also calculate the corresponding values of the vacuum energy density and the radial pressure measured by free-falling observers.

\subsection{Conformal factor at the horizon}
In order to obtain the values of the RSET in the $in$ vacuum, we need to calculate the conformal factor $C(u_{in},v_{in})$ which allows us to write the part of the metric \eqref{21} restricted to the time-radius subspace, in the $u_{in},v_{in}$ null coordinates,
\begin{equation}
ds^2_{(2)}=-C(u_{in},v_{in})du_{in}dv_{in}.
\end{equation}
Just as a reminder, the $in$ vacuum state of the dimensionally reduced problem is defined by the plane waves in the asymptotic Minkowski region at past null infinity, which is entirely within the asymptotic region of the exterior Schwarzschild geometry if the shell never reaches the speed of light in the past. The ingoing modes, labelled by $v_{in}$, either fall directly into the singularity or are reflected at the origin $r_-=0$ and from there either escape before the formation of the horizon and reach future null infinity, or fall into the singularity. If they reach future null infinity, they can be labelled by the coordinate $u_{out}$, the value of which is a function of the previous label $v_{in}$. Any point in the geometry outside the event horizon (both on the exterior and interior of the shell) can be labelled by a pair $(u_{out}, v_{in})$. In the notation introduced in eqs. \eqref{1} and \eqref{3}, $v_{in}$ is simply $v_+$ and $u_{out}$ is $u_+$. The dispersion of the light rays between past and future null infinity is given by
\begin{equation}
\frac{du_{out}}{dv_{in}}=\left.\frac{du_+}{du_-}\frac{dv_-}{dv_+}\right|_{v_-=u_-}=\frac{g(u_-)}{h(u_-)},
\end{equation}
where we have made use of the relation $du_-=dv_-$ for the reflection of light rays at the origin, and where $u_-$ is a function of $v_{in}$ through the inverse of the integral of $h(u_-)$. Studying the values of $g(u_-)$ and $h(v_-)$, defined in \eqref{6}, from $-\infty$ until the formation of the horizon for different trajectories of collapse, one can see that $h$ is of order one throughout. On the other hand, $g$ always has a divergence at the horizon since $u_+$ reaches an infinite value while $u_-$ is still finite. The contrast in this behaviour implies that the approximation
\begin{equation}
\frac{dv_-}{dv_{in}}\simeq 1,
\end{equation}
that is, the approximation of considering our $v_{in}$ coordinate as the Minkowski $v_-$, captures the relevant physical effects produced in the dynamics around the formation of the horizon. It is easy to check that introducing a function $h$ which is different from 1, but of the same order of magnitude, would not change the general aspects of the results. This approximation, apart from simplifying the calculations which follow, also allows us to fix the trajectory of the shell only in an arbitrarily small region around the point of horizon-crossing.\par
From this point on we will drop the subscripts from the two null coordinates we will use for the most part: $v\equiv v_+$ and $u\equiv u_-$ (we will not use $u_{out}$ since it is divergent at the horizon). Also, we will mostly use the radial coordinate in the exterior region, so $r$ will always refer to $r_+$.\par
From equations \eqref{3} and \eqref{6} we see that the conformal factor of the dimensionally reduced geometry as a function of $u$ and $v$ is
\begin{equation}
C(u,v)=|f(r(u,v))|g(u).
\end{equation}
Since we are interested in calculating the RSET at the horizon, where large values might be expected for it, we must evaluate the above quantity and at least its first two derivatives there. A minor inconvenience in that process is the fact that the explicit form of $r(u,v)$ is not generally available, and numerical calculations cannot be relied upon either, since at the horizon $f$ is zero and $g$ diverges. To handle this difficulty, we will use an expansion for $r(u,v)$ around the line corresponding to the horizon, where $u=u_{\rm h}=const.$,
\begin{equation}\label{8q}
r(u,v)=q_0(v)+q_1(v)(u-u_{\rm h})+\frac{1}{2}q_2(v)(u-u_{\rm h})^2+\cdots,
\end{equation}
where $q_i$ is the $i$-th derivative of $r$ with respect to $u$ evaluated at $u_{\rm h}$, namely, $q_i=\partial^i r/\partial u^i|_{u=u_{\rm h}}$. In order to calculate the RSET components, we will need up to second derivatives of the conformal factor in $u$. To evaluate them we must use the expansion of $r(u,v)$ in $u$ up to third order, due to the $1/(u-u_{\rm h})$ divergence generally present in $g(u)$. This means that we need only $q_0,\dots,q_3$.\par
Let us now see how to calculate these coefficients. The lowest order one $q_0(v)$ is just the value of $r$ at the horizon, namely, the constant $r_{\rm s}=2M$. The rest of them can be obtained through the relations
\begin{equation}\label{6q}
\frac{\partial r}{\partial u}=-\frac{1}{2}g(u)f(r),\qquad \frac{\partial r}{\partial v}=\frac{1}{2}f(r),
\end{equation}
as we show in the following. The first of these equations evaluated at $u_{\rm h}$ gives $q_1(v)$, but its rhs is just as difficult to evaluate as the conformal factor itself. However, we can make use of the second equation to write the cross-derivative
\begin{equation}\label{7q}
\frac{\partial}{\partial{v}}\frac{\partial{r}}{\partial u}=-\frac{1}{2}g(u)f'(r)\frac{\partial r}{\partial v}=-\frac{1}{2}g(u)f'(r)\frac{1}{2}f(r)=\frac{1}{2}f'(r)\frac{\partial r}{\partial u}.
\end{equation}
Taking into account that $f'(r)$ evaluated at the horizon is just $1/r_{\rm s}$, the evaluation of this equation at $u_{\rm h}$ gives us a first order differential equation for $q_1(v)$, namely $q_1'(v)=q_1(v)/(2r_{\rm s})$. Using this method recursively allows us to write analogous equations for all the coefficients $q_i(v)$ in \eqref{8q}. For the ones relevant to our calculation of the RSET we obtain
\begin{equation}\label{17q}
\begin{split}
&q_1'(v)=\frac{1}{2r_{\rm s}}q_1(v),\\
&q_2'(v)=\frac{1}{2r_{\rm s}}q_2(v)-\frac{1}{r_{\rm s}^2}q_1^2(v),\\
&q_3'(v)=\frac{1}{2r_{\rm s}}q_3(v)-\frac{3}{r_{\rm s}^2}q_2(v)q_1(v)+\frac{3}{r_{\rm s}^3}q_1^3(v).
\end{split}
\end{equation}
Initial conditions for these equations can be found by fixing the zero of the $v$ coordinate at the point of horizon formation, and considering the relation $r_+=r_-$ at the surface of the shell.\par
For a shell which crosses the horizon with an approximately constant radial velocity as seen from the inside ($\alpha_-=dv_-/du_-\simeq const.$), from equations \eqref{2} and \eqref{5} we get the relation
\begin{equation}\label{28q}
r_-\simeq r_{\rm s}+\frac{\alpha_--1}{2}(u-u_{\rm h})
\end{equation}
at the shell surface, which gives us the initial conditions $q_1(0)=(\alpha_--1)/2$, $q_2(0)=0$ and $q_3(0)=0$ (these last two are approximate if $\alpha_-$ is only approximately constant, but the important aspects of our final results do not change if they have different values). We solve the above equations to get
\begin{equation}\label{29q}
\begin{split}
&q_1(v)=-\frac{1-\alpha_-}{2}e^{v/2r_{\rm s}},\\
&q_2(v)=\frac{(1-\alpha_-)^2}{2r_{\rm s}}e^{v/2r_{\rm s}}(1-e^{v/2r_{\rm s}}),\\
&q_3(v)=-\frac{3(1-\alpha_-)^3}{8r_{\rm s}^2}e^{v/2r_{\rm s}}(1-4e^{v/2r_{\rm s}}+3e^{v/r_{\rm s}}).
\end{split}
\end{equation}

\subsection{RSET evaluated at the horizon for the ``in" vacuum}
\begin{figure}
	\centering
	\includegraphics[scale=0.7]{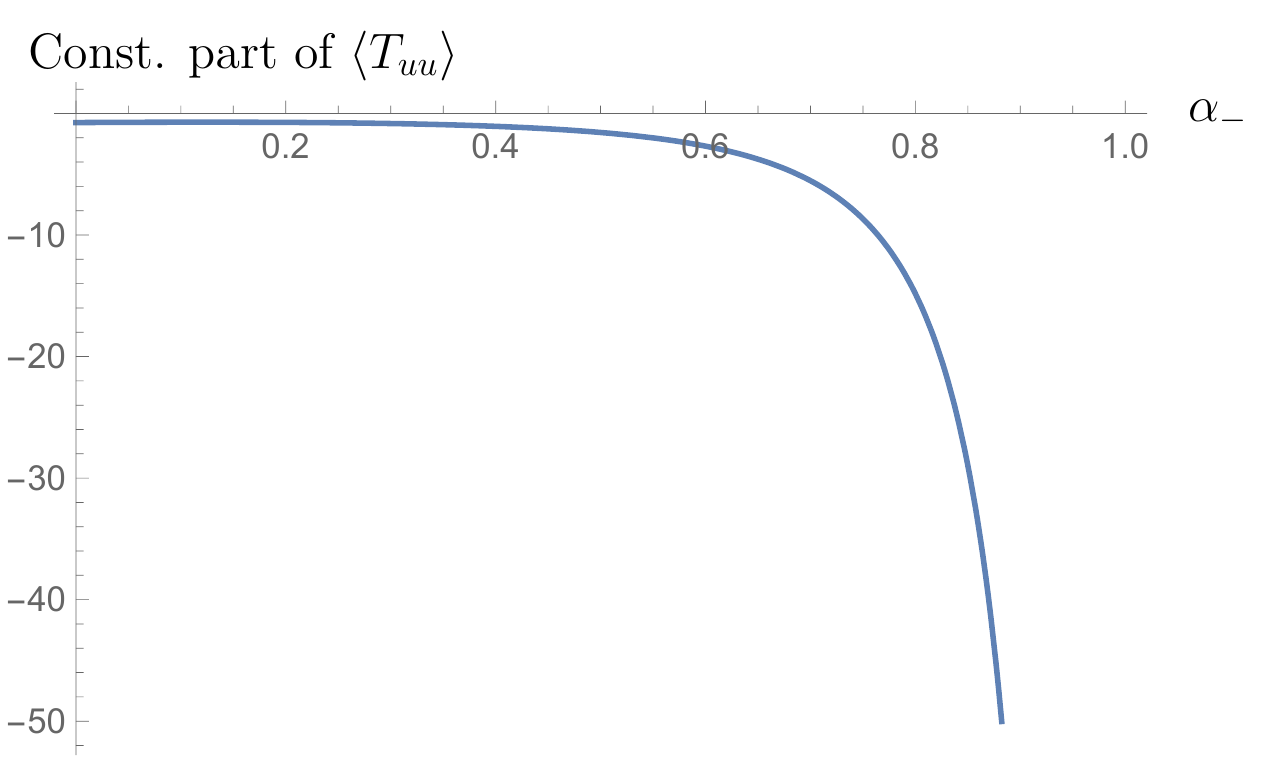}
	\caption{Plot of the constant part in the parenthesis of eq. \eqref{31a} as function of $\alpha_-$ in its domain of possible values. It has negative values throughout and a divergence at $\alpha_-=1$.}
	\label{f17}
\end{figure}

We now have everything prepared to calculate the RSET components at the horizon. Substituting the solutions \eqref{29q} into the series expansion \eqref{8q}, we see how $f$ depends on $u$ and $v$ up to third order in $(u-u_{\rm h})$. As for resolving the dependence of $g$ in $r$ (which appears through $\alpha_+$), we must remember the definition of this function \eqref{6} which tells us that it is evaluated at the shell surface. Therefore, close to the horizon, we can simply use the expression for $r$ given in \eqref{28q}. With these functions we can obtain $C(u,v)$ up to second order in $(u-u_{\rm h})$ (remember $g$ has a leading term $1/(u-u_{\rm h})$),
\begin{equation}
\begin{split}
C(u,v)&=(1-\alpha_-)e^{v/2r_{\rm s}}+\left[\left(-\alpha_-^2+\frac{3}{2}\alpha_--1\right)e^{v/2r_{\rm s}}+(1-\alpha_-)^2e^{v/r_{\rm s}}\right]\frac{u-u_{\rm h}}{r_{\rm s}}\\&+\frac{e^{v/2r_{\rm s}}}{8(1-\alpha_-)}\left[3-10\alpha_-+12\alpha_-^2-10\alpha_-^3+3\alpha_-^4\right.\\&\left.-4(1-\alpha_-)^2(3-5\alpha_-+3\alpha_-^2)e^{v/2r_{\rm s}}+9(1-\alpha_-)^4e^{v/r_{\rm s}}\right]\frac{(u-u_{\rm h})^2}{r_{\rm s}^2}+\cdots.
\end{split}
\end{equation}
Finally, we can use \eqref{23} to obtain the components of the RSET at the horizon:
\begin{subequations}\label{31q}
	\begin{align}
	\begin{split} \expval{T_{uu}}&=\frac{1}{24\pi r_{\rm s}^2}\left(\frac{-6\alpha_-^4+16\alpha_-^3-27\alpha_-^2-16\alpha_--6}{8(1-\alpha_-)^2}\right.\\&\hspace{35mm}\left.+\frac{\alpha_-}{2}e^{v/2r_{\rm s}}+\frac{3}{4}(1-\alpha_-)^2e^{v/r_{\rm s}}\right), \end{split}\label{31a}\\
	\expval{T_{uv}}&=-\frac{1}{24\pi r_{\rm s}^2}\frac{1-\alpha_-}{2}e^{v/2r_{\rm s}},\\
	\expval{T_{vv}}&=-\frac{1}{24\pi r_{\rm s}^2}\frac{1}{8}.
	\end{align}
\end{subequations}
Their behaviour can be read easily, except perhaps for the first constant term in the parenthesis in $\expval{T_{uu}}$, which has been plotted as a function of $\alpha_-$ in fig.~\ref{f17}. The following observations can be made:
\begin{itemize}
	\item Firstly, the components seem to grow exponentially on the horizon as time passes. This, however, turns out to be a consequence of the coordinate system in which they are expressed. In a system more appropriate for the static Schwarzschild region, say the Eddington-Finkelstein advanced coordinates $(v,r)$, this behaviour is suppressed by factors of $1/C$ arising from the relation $\partial u/\partial r$. A more detailed analysis of the energy density and flux perceived by a free-falling observer will follow shortly.
	\item The second thing one might notice is that $\expval{T_{vv}}$ is constant, and therefore completely independent from the dynamics of the collapse. This is an obvious consequence of the fact that we have chosen the Eddington-Finkelstein $v$ coordinate, which is not affected by the interior Minkowski region.
	\item Finally, we note that the $\expval{T_{uu}}$ component diverges as $\alpha_-\to 1$, that is, as the collapse becomes slower, approaching the static limit. As we will see, the $1/(1-\alpha_-)^n$ terms are suppressed exponentially in the regular Eddington-Finkelstein coordinates when a long time has passed since the formation of the horizon, but they play an important role near the point of horizon formation.
\end{itemize}

\subsection{Energy density, flux and pressure observed by free-falling observer at the horizon}
Let us consider the four-velocity $w$ of a free-falling observer in the Schwarzschild geometry expressed in $(u,v)$ coordinates, evaluated at the moment of horizon crossing. It has the form
\begin{equation}
w^\rho=\left(\sqrt{\frac{2}{\beta_0}}\frac{1}{C},\sqrt{\frac{\beta_0}{2}}\right),
\end{equation}
where $\beta_0$ is related to the radius $r_0$ from which the free fall was initiated through
\begin{equation}
\beta_0=\frac{1}{2}\frac{1}{1-r_{\rm s}/r_0}.
\end{equation}
Let us also introduce the space-like unitary vector perpendicular to this four-velocity and pointing in the outward radial direction,
\begin{equation}
z^\rho=\left(-\sqrt{\frac{2}{\beta_0}}\frac{1}{C},\sqrt{\frac{\beta_0}{2}}\right).
\end{equation}
We now define the effective energy density $\rho$, flux $\Phi$ and pressure $p$ perceived by this observer as
\begin{subequations}\label{36q}
	\begin{align}
	&\rho\equiv\expval{T_{\mu\nu}}w^\mu w^\nu=\frac{2}{\beta_0 C^2}\expval{T_{uu}}+\frac{1}{C}\expval{T_{uv}}+\frac{\beta_0}{2}\expval{T_{vv}},\\
	&\Phi\equiv-\expval{T_{\mu\nu}}w^\mu z^\nu=\frac{2}{\beta_0 C^2}\expval{T_{uu}}-\frac{\beta_0}{2}\expval{T_{vv}},\\
	&p\equiv\expval{T_{\mu\nu}}z^\mu z^\nu=\frac{2}{\beta_0 C^2}\expval{T_{uu}}-\frac{1}{C}\expval{T_{uv}}+\frac{\beta_0}{2}\expval{T_{vv}}.
	\end{align}
\end{subequations}
As an aside, we note that the conformal factor at the horizon,
\begin{equation}\label{37q}
C(u_{\rm h},v)=(1-\alpha_-)e^{v/2r_{\rm s}},
\end{equation}
is not equal to 1 when $v=0$, where the geometry must match with the interior Minkowski region, because we are not using the Minkowski $v_-$ coordinate. If we were, we would have to multiply $C$ by $h=dv_+/dv_-$, which at the point of horizon formation has the value $h=1/(1-\alpha_-)$).\par
\begin{figure}
	\centering
	\includegraphics[scale=0.55]{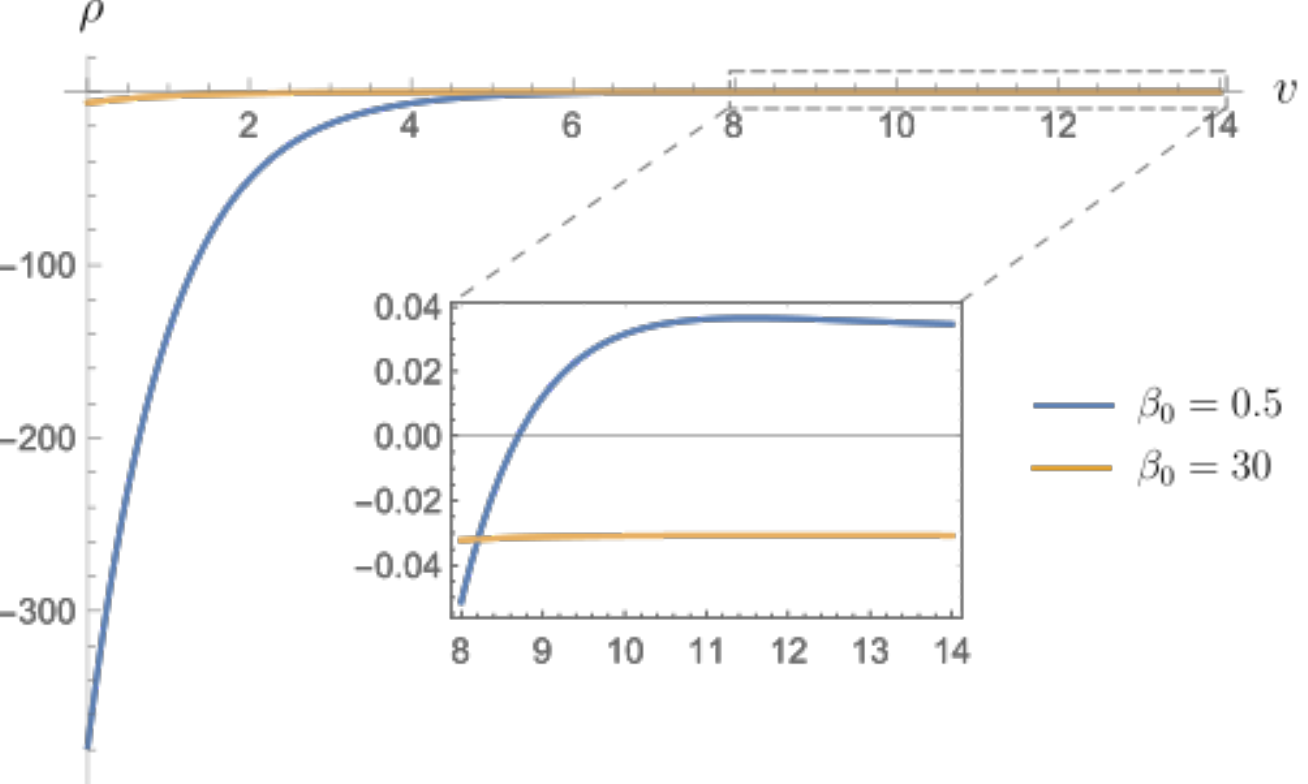}
	\includegraphics[scale=0.55]{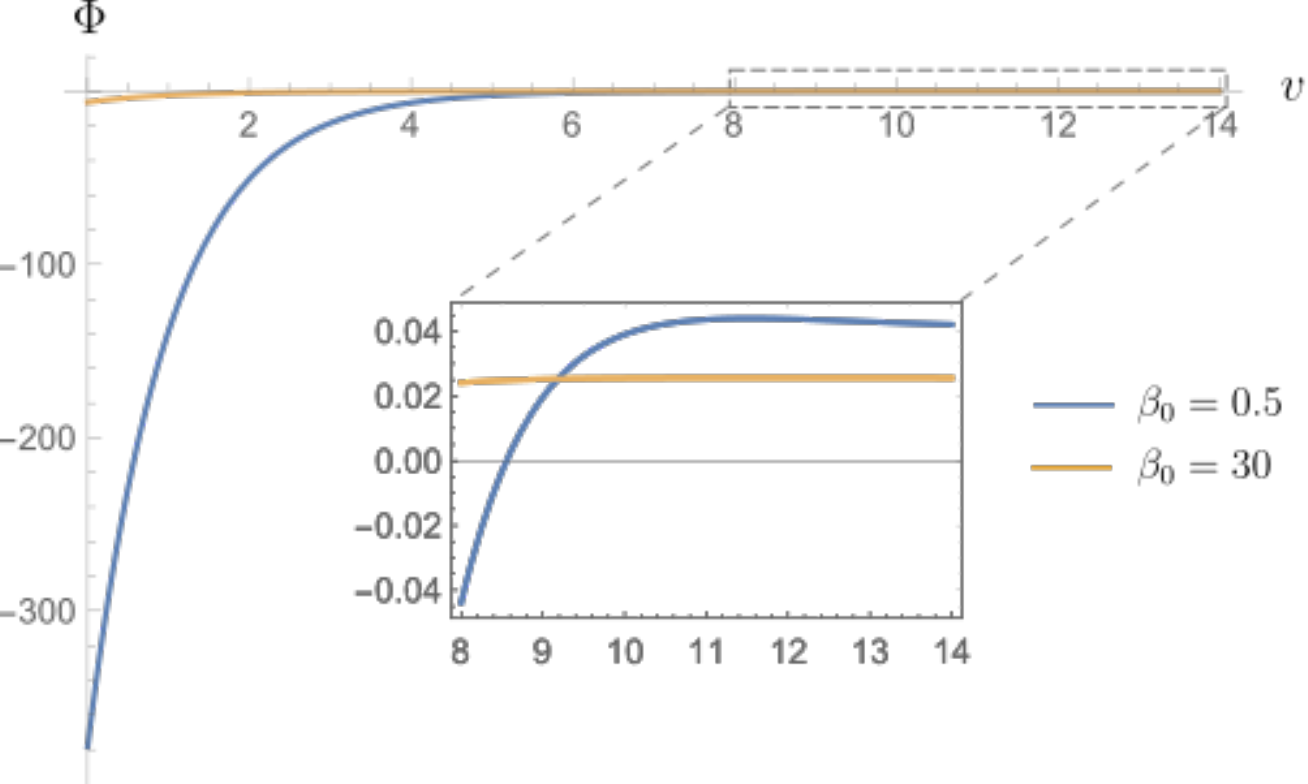}
	\includegraphics[scale=0.55]{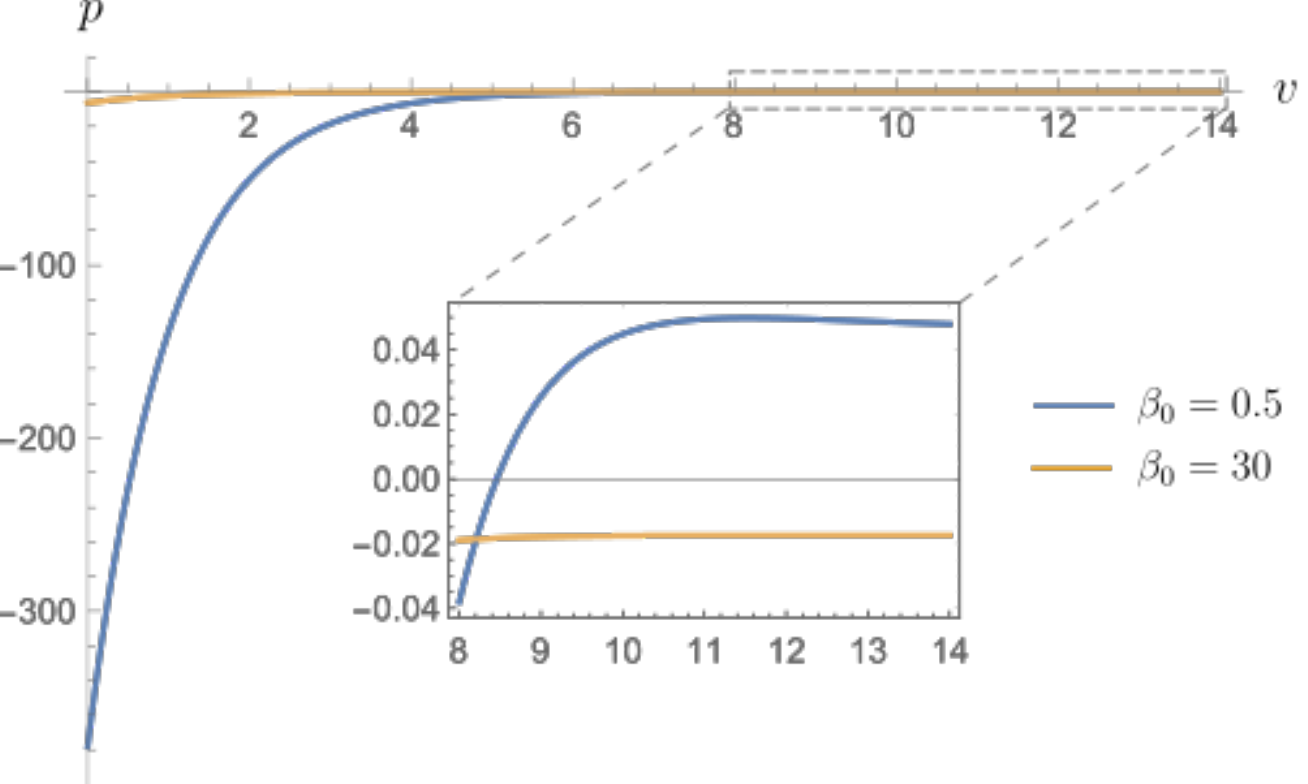}
	\caption{Perceived energy density, flux and pressure at the horizon, as a function of the Eddington-Finkelstein $v$ coordinate, for $\beta_0=0.5$ (free fall from $r_0\to\infty$) and for $\beta_0=30$ (free fall from $r_0\simeq1.017r_{\rm s}$), in the Schwarzschild region of a collapse with parameter $\alpha_-=0.9$ ($v_{\rm therm}\simeq9.2r_{\rm s}$). The point $v=0$ marks the formation of the horizon. Immediately after, we observe these function have very large (negative) values. This is a direct consequence of the proximity of the parameter $\alpha_-$ to 1, as discussed in the text. On the other hand, the asymptotic values are always small (except for observers which start their free fall very close to the horizon, where $\beta\to\infty$). The sign of the asymptotic energy density and pressure depend of the velocity of the observer (they are negative for slower observers), while the outgoing flux is always positive, in accordance with the evaporating black-hole scenario.}
	\label{f18}
\end{figure}

We thus see that the growing exponentials appearing in eqs. \eqref{31q} do not show up in the scalar quantities in \eqref{36q}. In fact, these turn out to have constant, finite asymptotic values that depend on the initial condition $\beta_0$ of the free falling observer
\begin{subequations}
	\begin{align}
	&\rho\xrightarrow[v\to\infty]{} \frac{1}{24\pi r_{\rm s}}\left(-\frac{1}{2}-\frac{\beta_0}{16}+\frac{3}{2\beta_0}\right),\\
	&\Phi\xrightarrow[v\to\infty]{} \frac{1}{24\pi r_{\rm s}}\left(\frac{\beta_0}{16}+\frac{3}{2\beta_0}\right),\\
	&p\xrightarrow[v\to\infty]{} \frac{1}{24\pi r_{\rm s}}\left(\frac{1}{2}-\frac{\beta_0}{16}+\frac{3}{2\beta_0}\right).
	\end{align}
\end{subequations}
When these values are approximately reached, the system can be said to have thermalised, as all other terms are suppressed exponentially. A measure of the time it takes to do so, in the $v$ coordinate, for a slow collapse (when $\alpha_-$ is close to 1) is given by the value
\begin{equation}
v_{\rm therm}= 4r_{\rm s}\log(\frac{1}{1-\alpha_-}).
\end{equation}
In fig.~\ref{f18} we see plots for $\rho$, $\Phi$ and $p$ for two different values of $\beta_0$, which make the asymptotic values of $\rho$ and $p$ have different signs. Except for the case of extremely small values of $\beta_0$, the asymptotic values of the previous quantities are always negligibly small due to the suppression of the RSET by Planck's constant (which has been omitted in the choice of units). However, this smallness can be compensated during the transient phase of the collapse. Near the point of horizon formation we have $1/(1-\alpha_-)^4$ terms, originating from the $1/(1-\alpha_-)^2$ term in $\expval{T_{uu}}$ in \eqref{31q} and from the $1/C^2$ term evaluated from \eqref{37q}. These terms can be made arbitrarily large if $\alpha_-$ is very close to 1, compensating the suppression by Planck's constant.\par

With these results we see that the RSET approaches a physical divergence in the static limit $\alpha_-\to1$. At the limit itself this divergence is hardly surprising, as for a static shell the $in$ vacuum essentially becomes the Boulware one. What is interesting is the fact that this limit can be approached through the single velocity parameter $\alpha_-$ of the shell when it crosses the horizon, without imposing any conditions on its past evolution. This seems to indicate that during the formation of a black hole, if by some mechanism the collapse of matter were to be slowed down just before it forms a horizon, its subsequent evolution would become a problem which requires a full semiclassical treatment.

\section{Summary and conclusions}

In the standard picture of a black-hole formation process, the pressure of a star fails to support its structure and matter begins to accelerate in an inward direction, acquiring very high speeds by the time a horizon is formed. In this scenario, the semiclassical theory presents no significant deviations from its classical counterpart \cite{DFU,Parentani1994,Barcelo2008,Unruh2018}. This is true throughout the collapse, except perhaps when the curvature approaches Planckian values, in the final stages before a singularity is formed, although it is not clear whether the semiclassical theory is applicable there at all. However, whether semiclassical effects become important in scenarios involving matter approaching the formation of a horizon in a different manner is less well understood, and worth studying in detail in order to determine the possible self-consistency of models of gravitational collapse beyond general relativity.\par
Our study is inspired by the possible behaviour of matter in such situations, covering nonetheless a very large family of geometries due to the many unknowns in their evolution. In section \ref{s4} we begun by analysing the case of an oscillating distribution of matter in the thin-shell approximation, which periodically approaches the formation of a horizon but bounces back just before it is formed. Through the ETF we saw periods of emission of Hawking-like radiation in between the bounces. Although the ETF in these periods was always around the Hawking value $1/2r_{\rm s}$, its particular shape was strongly influenced by how the in-crossing and out-crossing dispersion effects resonated with each other for individual modes. The bounces themselves caused a significant dispersion in the out-crossing modes, which translates into sharp increases of both the ETF and RSET. In general, we saw that semiclassical quantities (which depend on the derivatives of the terms measuring light-ray dispersion) become largest near the Schwarzschild radius at low speeds.\par
To further explore this low-velocity regime, in section \ref{s5} we analysed surface trajectories which approach the Schwarzschild radius monotonously, but reach it only asymptotically. In this case we saw that semiclassical effects are very sensitive to the structure of the geometry close to the surface, so we needed to go beyond the thin-shell approximation and use an arbitrary spherically-symmetric geometry for the interior. With minimal assumptions, we showed that the values of the ETF and RSET at large times depend only on a few characteristics of the geometry through one of its degrees of freedom, which we called the generalised redshift function, particularly: the speed at which its minimum approaches zero (i.e. the speed at which the formation of an apparent horizon is approached) and its spatial derivatives on both sides of this minimum. Depending on these quantities, the dynamical $in$ vacuum can behave as in the usual case of black-hole formation in finite time, or it can become similar to the static Boulware vacuum (generally at lower speeds of approach). In the latter case the RSET acquires very large values around the Schwarzschild radius, tending to a divergence asymptotically.\par
In section \ref{s6} we went back to the thin-shell approximation, and analysed the case of a trajectory which forms a horizon in finite time. The parameter we were interested in was the speed at which the shell crossed the Schwarzschild radius. We calculated the values of the RSET at the horizon, and the corresponding energy density, flux and pressure perceived by free-falling observers, with a dependence on this speed parameter. We saw that at low speeds these physical quantities can become arbitrarily large (and also stay large for longer at lower speeds), approaching a divergence in the static limit.\par
To conclude, we remark that a clear-cut result from all the above situations is that semiclassical back-reaction on the geometry (through the RSET) is a necessary ingredient in analysing any geometry in which matter happens to be moving at very low velocities (much lower than the speed of light) when close to horizon formation. As a purely kinematic exercise, our analysis shows the richness of the situations around the threshold of horizon formation. Beyond that, although no complete dynamical scenario has yet been developed in which matter actually enters such low-velocity regimes, it is important to note that such a possibility is not excluded either. This offers the exploration of alternative scenarios with which to compare the standard black-hole paradigm.

\section*{Acknowledgments}
Financial support was provided by the Spanish Government through the projects FIS2017-86497-C2-1-P, FIS2017-86497-C2-2-P (with FEDER contribution), FIS2016-78859-P (AEI/FEDER,UE), and by the Junta de Andalucía through the project FQM219. VB is funded by the Spanish Government fellowship FPU17/04471.

\section*{References}

\nocite{*}
\bibliography{Bibliografia}
\bibliographystyle{ieeetr}

\end{document}